\RequirePackage{fix-cm}
\documentclass[smallextended]{svjour3_B}       
\usepackage{graphicx}
\usepackage{url}
\usepackage{booktabs}
\usepackage{multirow}
\usepackage{hyperref}
\usepackage{amssymb}
\usepackage{datetime}

\begin{document}

\title{{Engineering cryogenic setups for 100-qubit scale superconducting circuit systems}}

\subtitle{}


\author{S.~Krinner, S.~Storz, P.~Kurpiers, P.~Magnard, J.~Heinsoo, R.~Keller, J.~L\"utolf, C.~ Eichler, A.~Wallraff}

\authorrunning{Krinner et al.} 

\institute{Department of Physics, ETH Z\"urich, CH-8093, Z\"urich, Switzerland}

\date{Dated: June 20, 2018}

\maketitle

\begin{abstract}
A robust cryogenic infrastructure in form of a wired, thermally optimized dilution refrigerator (DR) is essential for present and future solid-state based quantum processors. Here, we engineer an extensible cryogenic setup, which minimizes passive and active heat loads, while guaranteeing rapid qubit control and readout. We review design criteria for qubit drive lines, flux lines, and output lines used in typical experiments with superconducting circuits and describe each type of line in detail. The passive heat load of stainless steel and NbTi coaxial cables and the active load due to signal dissipation are measured, validating our robust and extensible concept for thermal anchoring of attenuators, cables, and other microwave components.
Our results are important for managing the heat budget of future large-scale quantum computers based on superconducting circuits.

\end{abstract}

\section{Introduction}
Promising solid-state based quantum computing platforms, such as superconducting circuits \cite{devoret_superconducting_2013} or charges and spins in semiconductor quantum dots \cite{awschalom_quantum_2013}, require temperatures on the Millikelvin level to initialise the systems in their ground state and to avoid errors due to thermal excitation during operation. Millikelvin temperatures are achieved in ${\rm He^3/He^4}$ dilution refrigerators (DR). When scaling those approaches from the few qubit level to large scale quantum processors, an increasing number of microwave and DC cables need to be integrated into the dilution refrigerator. They connect the classical control electronics at room temperature (RT) to the quantum processor at the lowest temperature stage of the dilution refrigerator, creating a substantial heat load on the dilution refrigerator due to heat conduction. Besides this passive load, active load due to the dissipation of control signals in cables and attenuators plays a major role. Dissipation is necessary to thermalize the incoming radiation fields and to reduce the number of thermal photons incident on the sample \cite{yan_flux_2016,yeh_microwave_2017}.

While superconducting quantum processors operating 20 qubits have been recently realized \cite{IBMQ,RigettiQ}, their cryogenic setup and extensibility towards larger system sizes has not been reported on. Here, we present a thermally optimized, robust cabling scheme and cryogenic setup suitable for the operation of 50 qubits at a temperature of 14\,mK. Disregarding space constraints in the dilution refrigerator, at least 150 qubits could be operated. Before describing our customized cryogenic system in Section \ref{sec:setup}, we put forward methods of how to minimize passive heat loads on the stages of the dilution refrigerator and how to minimize thermal photons in cables connecting the room temperature to the base temperature stage while keeping active load low (Section \ref{sec:diffheatloads}).
We then present measurements of the passive loads arising from subsequent installations of cable trees comprising typically 25 coaxial lines each, and compare them to estimates based on available data on thermal conductances of coaxial cables (Section \ref{sec:passiveloadmeas}). Active loads due to application of control signals are characterized in Section \ref{sec:activeloadmeas}. Finally, we discuss the total heat load in Section \ref{sec:conclusion} along with possible improvements which would allow to increase the number of operational qubits to up to thousand when disregarding space constraints.

\subsection{Typical cabling for experiments with superconducting circuits}
When designing the control lines and output lines connecting to a superconducting quantum processor (QP) the goal is to provide sufficiently strong coupling rates to the quantum processor, while minimizing decoherence due to coupling of the QP via these lines to its environment.
Thermal noise, present due to the connection of the QP to electronics at room temperature, not only leads to qubit dephasing \cite{ithier_decoherence_2005,yan_flux_2016,yeh_microwave_2017}, but can also lead to creation of quasi-particles and thus to dissipation and reduced energy relaxation times \cite{catelani_quasiparticle_2011,barends_minimizing_2011,corcoles_protecting_2011,pop_coherent_2014,kreikebaum_optimization_2016}. Hence, thorough thermalization of cables, attenuators, and microwave components at the various temperature stages of the dilution refrigerator is not only important for reducing the heat load on the dilution refrigerator, but also for protecting the QP from thermal radiation. In addition to thermal anchoring, filters with stop-bands outside the frequency range of qubits and readout resonators as well as infra-red blocking filters \cite{lukashenko_improved_2008,Kurtis_phd_2013,kreikebaum_optimization_2016} further suppress thermal radiation. Other sources of external noise that can negatively impact coherence times include 1/f noise \cite{ithier_decoherence_2005,koch_charge-insensitive_2007} from electronic instruments and magnetic field fluctuations or magnetic offset fields inducing magnetic vortices that can lead to dissipation \cite{song_microwave_2009,nsanzineza_trapping_2014,kreikebaum_optimization_2016}. These sources can be mitigated with appropriate filters and magnetic shielding respectively, but their study is not a focus of this work.


We briefly present an overview of the cabling typically used in experiments with superconducting circuits. We distinguish between direct-current (DC) and radio-frequency (RF) cabling. DC lines are made from twisted pair wires, that are typically low-pass filtered and thermalized at each temperature stage. Typical scopes of applications are biasing of cryogenic amplifiers and flux biasing of tunable frequency qubits \cite{hutchings_tunable_2017}. RF lines on the other hand are realized as semi-rigid microwave cables and contain various microwave components such as attenuators, filters or amplifiers. They connect to the QP and are used for control and readout of the quantum processor. One typically distinguishes between drive lines, flux lines, and output lines, which we briefly described here.

Drive lines are used for driving qubits with a microwave tone realizing single-qubit gates, and for driving readout resonators. To reduce thermal population of qubits, and frequency shifts of the qubits due to their dispersive interaction with a readout resonator \cite{blais_cavity_2004}, the number of thermal noise photons in the drive lines arriving at the mixing chamber plate (MXC) of the dilution refrigerator should be well below the single photon level in both cases. More precisely, to guarantee a noise photon number at MXC on the $10^{-3}$ level, a total attenuation of about 60\,dB is required, see Section \ref{sec:attenuation} and \ref{sec:setupdrivelines}. The bandwidth of the drive lines is required to be large enough to cover the typical frequency ranges of qubits (4-6\,GHz) and of readout resonators (4-8\,GHz).

Flux lines are mainly used for implementing two-qubit gates which are based on the dynamical tunability of the transition frequency of a qubit \cite{dicarlo_demonstration_2009,barends_superconducting_2014,caldwell_parametrically_2017} or of a separate coupling subcircuit \cite{mckay_universal_2016}. In addition, qubit frequency variations, occurring due to imperfections in the fabrication of Josephson junctions, can be compensated. Tunable frequency qubits, as opposed to fixed frequency qubits, make use of a SQUID loop instead of a single Josephson junction as the inductive element. By threading a magnetic flux $\Phi$ through this loop the frequency of the qubit approximately changes as $\omega_q\simeq\omega_0 \sqrt{|{\rm cos}(\pi\Phi/\Phi_0)|}$ \cite{koch_charge-insensitive_2007}, where $\Phi_0$ is the magnetic flux quantum.
A mutual inductance between flux line and SQUID loop is realized by routing the on-chip part of the flux line past the SQUID loop. Hence, a current applied to the flux line results in a magnetic flux in the SQUID loop, effectively tuning the transition frequency of the qubit. Low-pass filters in the flux lines limit the bandwidth to about 1\,GHz eliminating thermal noise at qubit frequencies. However, since the magnetic flux $\Phi$ in the SQUID loop sets the qubit transition frequency, current noise leads to qubit dephasing \cite{ithier_decoherence_2005,koch_charge-insensitive_2007}. To reduce it, a suitable amount of attenuation (10-20\,dB) is also added in the flux lines, see Section \ref{sec:setupfluxlines}.

Output lines contain a series of cryogenic and room temperature amplifiers for the sensitive detection of readout signals \cite{wallraff_approaching_2005,vijay_observation_2011,walter_rapid_2017}. To isolate the sample from thermal noise photons and from noise of amplifiers while not attenuating the output signal, isolators and circulators are used, see Section \ref{sec:setupoutputlines} for more details.




\section{Sources of heat loads}\label{sec:diffheatloads}
We consider three dominant contributions to the heat load on the dilution refrigerator. First, passive load is due to heat flow from higher temperature stages to lower temperature stages. Here, we consider only heat conducted through installed cables. Heat which flows via posts that separate the various plates of the dilution refrigerator is not considered because it is already taken into account in the specified cooling power of the dilution refrigerator \cite{Leiden,Oxford,Bluefors}. Second, active load arises due to the dissipation (Joule heating) of applied microwave signals in attenuators and in the microwave cables themselves. Dissipation arising from DC signals, used e.g. to bias HEMT amplifiers at the 4K stage or to flux bias qubits on the chip at MXC, falls also in this category. Third, radiative load from stages and shields of higher temperature to stages and shields of lower temperature \cite{parma_cryostat_2015}. This load, however, is also taken into account in the specified cooling power of the dilution refrigerator. For completeness, we mention load due to residual gas, in particular Helium since it has the highest vapor pressure at cryogenic temperatures. However, during normal operation of the dilution refrigerator, the cryo-pumping capacity of the cold surfaces in the dilution refrigerator keeps the pressure in the vacuum can below $10^{-5}$\,mbar providing an adequate isolation vacuum \cite{parma_cryostat_2015}. We therefore do not consider this load.

\subsection{Passive load}
To minimize passive heat load we use cable materials with low thermal conductivity. Note that with the exception of superconductors this typically goes along with poor electrical conductivity, and hence leads to more dissipation when microwave signals are applied. However, in most of the lines attenuation and thus dissipation is anyway desired in order to thermalize the incoming radiation fields, see Section \ref{sec:activeload}.

To estimate the passive load due to installed cables we consider fixed temperatures on the plates of the dilution refrigerator, corresponding to a steady state, in which the heat flows onto the various stages are absorbed by the cooling powers available at those stages. Note that the cooling power on a given stage depends on the temperature of that stage. The heat flow $P_i$ to stage $i$ with temperature $T_i$ due to a single coaxial cable is given by
\begin{equation} \label{eqn:heat-flow}
P_i = \int_{T_{i-1}}^{T_i}dT\,\frac{\rho_{\rm o}(T) A_{\rm o} + \rho_{\rm d}(T) A_{\rm d} + \rho_{\rm c}(T) A_{\rm c}}{L_{i}},
\end{equation}
where $\rho_{\rm o}$, $\rho_{\rm d}$, $\rho_{\rm c}$, and $A_{\rm o}$, $A_{\rm d}$, $A_{\rm c}$ are thermal conductivities and cross sections of outer conductor (OC), dielectric, and center conductor (CC) respectively, and $L_i$ is the length of the cable connecting stage $i-1$ and $i$. $P_i$ has units of power. The labeling $i\in\{1,2,3,4,5\}$ corresponds to the stages \{"50K", "4K", "Still", Cold plate ("CP"), "MXC"\}. Table \ref{tab:DR} lists the corresponding temperatures in our Bluefors XLD400 dilution refrigerator as achieved in the empty dilution refrigerator after installation. The values are typical for state of the art pulse tube cooler based dilution refrigerators \cite{uhlig_3he/4he_2002,uhlig_dry_2008}. The thermal conductivities used for the estimates in this paper are listed in Appendix \ref{app:thcond}.

\renewcommand{\arraystretch}{2.1}
\begin{table}
\centering
\begin{tabular}{cccc}
\hline
\textbf{Stage name} & \textbf{Temperature (K)} & \textbf{Cooling power (W)} & \textbf{Cable length (mm)} \\
\hline
50K & 35 & 30 (at 45\,K) & 200 \\
4K & 2.85 & 1.5 (at 4.2\,K) & 290 \\
Still  & $882\times 10^{-3}$ & 40\,$\times 10^{-3}$ (at 1.2\,K) & 250  \\
CP  & $82\times 10^{-3}$ & 200$\times 10^{-6}$ (at 140\,mK) & 170  \\
MXC & $6\times 10^{-3}$ & 19$\times 10^{-6}$ (at 20\,mK)& 140  \\
\hline
\end{tabular}
\caption{{\bf Dilution refrigerator specifications.} Temperatures and available cooling powers on the indicated stages of a Bluefors XLD400 DR. Coaxial cable lengths towards the respective stages are listed as well.}
\label{tab:DR}
\end{table}

\begin{figure}
	\center
	\includegraphics{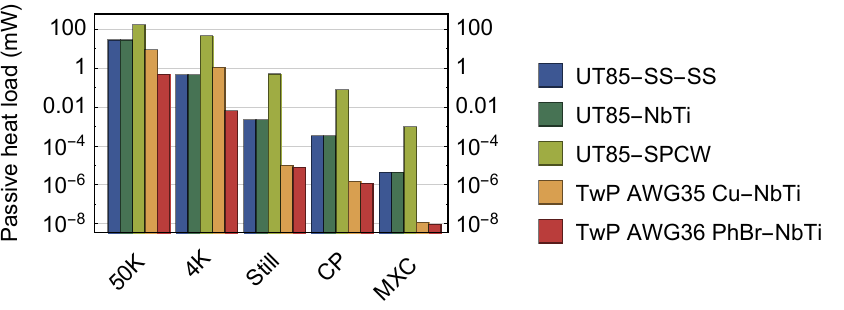}
	\caption{{\bf Heat flow for indicated cable types.} Heat flow to the five temperature stages for indicated cable types and given cable lengths (see text).
Twisted pairs (TwP) consist of either Cu or PhBr from RT to the 4K stage, and of NbTi from the 4K to the MXC stage.}
	\label{fig:heatFlow-materials}
\end{figure}

To motivate the choice of cable materials detailed in Section \ref{sec:setup} we plot in Fig. \ref{fig:heatFlow-materials} the heat flow to the various temperature stages due to single coaxial cables of different material.
Note that these estimates assume thermalization of outer conductor, dielectric, and center conductor at every temperature stage, see Section \ref{sec:detPassiveLoads} for a discussion.
Three typical and readily available types of coaxial cable are shown. The lengths of the cables are listed in Table \ref{tab:DR} and correspond to the inter-plate distances in the dilution refrigerator plus a small correction due to the bends in those cables (see Section \ref{sec:setup}). Stainless steel cables (UT85-SS-SS) and NbTi cables (UT85-NbTi) have the lowest passive load. This is why we choose them for drive and output lines, respectively. The flows of these two cable types are dominated by their outer conductor, which has a cross-sectional area that is by a factor 10 larger than the center conductor. The contribution of the Teflon dielectric is of the same order of magnitude as the one of the inner conductor. This results from the low thermal conductivity of Teflon, see Appendix \ref{app:thcond}, and the cross-sectional area of the dielectric being comparable to the one of the outer conductor.

An alternative to stainless steel cable is Cupronickel (CuNi) cable. It is expected to have a 20-30\% higher passive load than SS-SS cable. Another commonly used cable type is UT85-SS cable (stainless steel CC, SPCW\footnote{We model the SPCW center conductor as a single core Cu wire with an RRR of 100, which agrees well with measurements.} center conductor). It has has a larger load in general, in particular on the 4K stage due to the anomalously large thermal conductivity of Copper in the temperature interval 4-50\,K, see Appendix \ref{app:thcond}. Due to its comparably low attenuation, see Appendix \ref{app:cableattenuation}, it is suited for the sections in the output lines, for which NbTi is not superconducting, i.e. from the vacuum flange of the dilution refrigerator to the 4K stage, see also the discussion in Section \ref{sec:setupoutputlines}.

In our dilution refrigerator DC wiring is pre-installed. Copper or Phosphor-Bronze (PhBr) twisted pairs of diameters AWG35 and AWG36 respectively are only used from room temperature to the 4K stage due to their large thermal conductivity. From 4K to MXC superconducting NbTi twisted pairs are used\footnote{For clarity, NbTi twisted pairs are considered to consist of pure NbTi, although they are typically embedded in a CuNi matrix for mechanical stability.}. Fig. \ref{fig:heatFlow-materials} shows that among these materials only Cu twisted pairs cause a significant passive load comparable to the one of the coaxial cables.

\subsection{Active load}\label{sec:activeload}
The active load in the dilution refrigerator depends on the attenuation installed in the RF lines and on the signal levels required at the chip. We therefore in this section briefly discuss how much attenuation is needed to suppress thermal noise and how to distribute the attenuators among the various temperature stages. Furthermore, we discuss the signal levels required in typical experiments with superconducting qubits.

\subsubsection{The need of attenuation}\label{sec:attenuation}
Although the signals required to drive qubits or to read them out are small (e.g. a peak power of -66\,dBm for a 30\,ns long $\pi$-pulse), the power applied at the input of the dilution refrigerator needs to be orders of magnitude larger. This is because a total attenuation of at least $\sim 60\,$dB is required to reduce room temperature blackbody radiation, present in cables thermalized at room temperature, to a level corresponding to a thermal photon occupation number of a few $10^{-3}$ at the sample. This one-dimensional blackbody radiation in cables is known as Johnson-Nyquist noise in electronics. If a resistor $R$ is connected to the cable or if the cable is coaxial with a characteristic impedance $R$ the induced voltage fluctuations have a two-sided power spectral density given by
\begin{equation}\label{eqn:JNnoisequantum}
S_{V}^{\rm th}(T) = 2 {\rm k_B} T R \frac{h|\nu|}{{\rm k_B} T} \frac{1}{\textup{exp}(h|\nu|/{\rm k_B} T)-1}.
\end{equation}
The last factor in this equation is the photon occupation number as dictated by the Bose-Einstein distribution $n_{\textup{BE}}=1/\left[ \textup{exp}(h|\nu|/{\rm k_B} T)-1 \right]$. It is dimensionless and can be thought of as number of photons per Hz frequency interval, per second. In the classical limit $h\nu\ll{\rm k_B} T$ one recovers the original (two-sided) Johnson Nyquist formula $S_{V}^{\rm th}(T) = 2 {\rm k_B} T R$.

In a coaxial cable connecting room temperature to base temperature, thermal photons
propagate down the line towards the lower temperature stages. To reduce the spectral density of this radiation a series of attenuators is installed in the microwave line.
An attenuator with an attenuation of $A=20\,\textup{dB}=100$  can be considered to effectively function as a beamsplitter which transmits 1\% of the incident signal and adds 99\% of blackbody radiation with the effective temperature $T_{\rm att}$ at which the attenuator is thermalized. We note that 99\% of the incident signal is dissipated in the attenuator. To prevent the attenuator from heating up and to keep the effective temperature at which it re-emits low it is efficiently thermalized. More formally, the noise photon occupation number $n_i$ at stage $i$ with attenuation $A_i$ is given by
\begin{equation} \label{eqn:attenuator}
n_i(\nu)=\frac{n_{i-1}(\nu)}{A_i} + \frac{A_i-1}{A_i}n_{\textup{BE}}(T_{i,{\rm att}},\nu).
\end{equation}
When installing a cascade of attenuators in the microwave line thermalized at subsequently lower temperature stages, we employ this relation to determine the noise photon occupation number at the MXC stage, $n_{\rm MXC}$.

For our considerations we fix the signal frequency to $\nu_0=6$\,GHz since typical qubits are operated at 5\,GHz and typical readout resonators at 7\,GHz. A lower bound for the total attenuation needed to achieve a noise photon number of $n_{\rm MXC}=10^{-3}$ is obtained by neglecting the second term on the r.h.s. of Eqn. \ref{eqn:attenuator} yielding $n_{\textup{BE}}(T=300\,{\rm K},\nu_0)/10^{-3}=60$\,dB. This is a lower bound since blackbody radiation emitted by attenuators at all other temperature stages is neglected. 

\begin{figure}
	\center
	\includegraphics[width=0.45\textwidth]{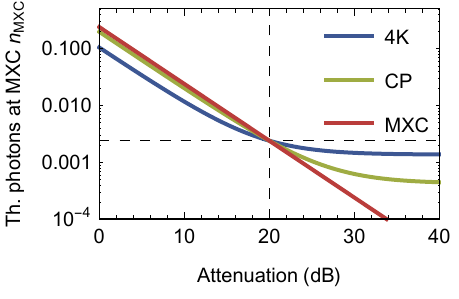}
	\hspace{0.2cm}
	\includegraphics[width=0.45\textwidth]{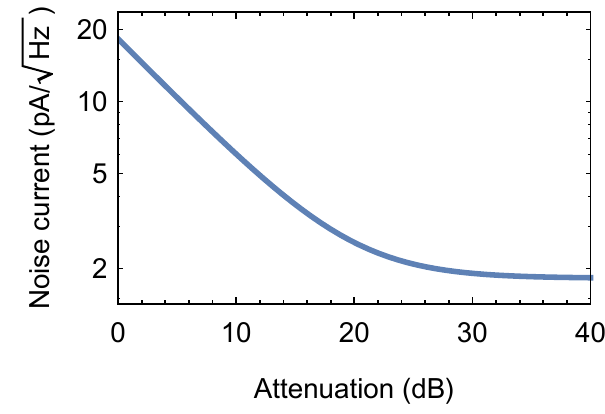}
	\caption{{\bf Thermal noise photon numbers. }a) Calculated thermal photon number at MXC as a function of the degree of attenuation at the 4K stage (blue), CP (green), MXC (red) for a fixed attenuation of 20dB at the respective other stages (45K stage and 1K stage have no attenuators in this example). The calculations are based on Eqn. \ref{eqn:attenuator}. b) Thermal noise current spectral density at MXC as a function of attenuation at the 4K stage, calculated using Eqn. \ref{eqn:currentnoise}.}
	\label{fig:thermal-photons}
\end{figure}

Distributing the total attenuation among the various temperature stages is important to keep the active load on the lower stages well below the cooling powers available at those stages. In distributing the attenuation we avoid putting more attenuation $A_i$ on stage $i$ than is needed to thermalize the incoming radiation fields onto that stage, i.e. the first term on the r.h.s. of Eqn. \ref{eqn:attenuator} should not be significantly smaller than the value to which the second term saturates for $A_i\gg1$. Hence, good reference values for the attenuation on stage $i$ are $A_{i,{\rm ref}}=n_{\textup{BE}}(T_{i-1},\nu_0)/n_{\textup{BE}}(T_{i},\nu_0)$. As an example we consider a total attenuation of 60\,dB that is composed of 20\,dB attenuators on the 4K, CP, and MXC stages. We first plot $n_{\rm MXC}$ as a function of $A_{\rm 4K}$ for a fixed attenuation of 20\,dB at the CP and MXC stages respectively, see blue solid line in Fig. \ref{fig:thermal-photons}. We observe that the number of noise photons is efficiently reduced by increasing the attenuation up to $\sim20\,$dB, as expected from the reference value $A_{\rm 4K, guess}=n_{\textup{BE}}(T_{\rm RT},\nu_0)/n_{\textup{BE}}(T_{\rm 4K},\nu_0)\approx T_{\rm RT}/T_{\rm 4K}=300/3=20$\,dB. More attenuation is unnecessary because at this value the radiation field incident from room temperature has reached the 4K thermal noise floor.

A similar property is found when plotting $n_{\rm MXC}$ as a function of $A_{\rm CP}$, for a fixed attenuation of 20\,dB at the 4K and MXC stages respectively (green dashed line in Fig. \ref{fig:thermal-photons}). Here a little bit more than 20\,dB attenuation would further reduce $n_{\rm MXC}$, as expected from a reference value $A_{\rm CP, ref}=n_{\textup{BE}}(T_{\rm 4K},\nu_0)/n_{\textup{BE}}(T_{\rm CP},\nu_0)=22$\,dB.
We finally plot $n_{\rm MXC}$ as a function of attenuation at MXC (red dash-dotted line in Fig. \ref{fig:thermal-photons}). It demonstrates that over the considered attenuation range adding more attenuation at MXC linearly decreases $n_{\rm MXC}$. This is because the noise floor of thermal photons at MXC is as low as $n_{\textup{BE}}(T_{\rm MXC}=20\,{\rm mK},\nu_0)=6\times 10^{-7}$. However, we will see in section \ref{sec:setupdrivelines} that the limited cooling power at MXC prevents us from installing significantly more than 20\,dB of attenuation at MXC.

\subsubsection{Signal levels required for the operation of the QP}\label{sec:requsignallevel}
In this section we estimate the powers required at the chip for two important operations on superconducting qubits: driving a $\pi$-pulse on a qubit and setting a flux bias on a qubit.

To drive a $\pi$-pulse on a qubit, we apply an RF pulse at the qubit frequency with a Gaussian envelope through a CPW transmission line weakly capacitively coupled to the qubit. This coupling has to be sufficiently small to prevent Purcell decay of the qubit into the transmission line \cite{Pechal2016a}. The finite value of the coupling imposes a limit on the $T_1$ time of the qubit, which we target here to be no lower than $T_1^{\rm lim}=0.5\,$ms, significantly larger than state-of-the-art coherence times \cite{sheldon_characterizing_2016,yan_flux_2016}. The amplitude of the time-dependent Rabi frequency $\Omega_R(t) = \Omega_0 {\rm exp}\left[-t^2/(2\sigma^2)\right]$ for a 30\,ns long $\pi$-pulse with a width of $\sigma=5\,$ns is given by $\Omega_0=\sqrt{\pi/(2\sigma^2)}\approx2\pi\times 40$\,MHz. To achieve this value a peak power $P_{\rm p}=\hbar\omega_q T_1^{\rm lim}\Omega_0^2/4\approx -66\,$dBm is needed \cite{Pechal2016a}. For an estimate of the associated heat load we employ the average power of the pulse, $P_{\rm avg}=P_{\rm p}\frac{1}{6\sigma}\int_{-3\sigma}^{3\sigma}{\rm exp}\left[-t^2/(2\sigma^2)\right]\,{\rm d}t=\frac{\sqrt{\pi}}{6}P_{\rm p}\approx -71\,$dBm. This number will be further reduced due to a finite duty cycle of those pulses during the execution of a quantum algorithm. We assume a maximum duty cycle of 33\,\%, corresponding to an operation mode of the quantum processor where single- and two-qubit gates are alternated with the two-qubit gate duration being twice as long as the single-qubit gate duration \cite{barends_superconducting_2014,martinis_fast_2014}. Also $\pi/2$-pulses require only a quarter of the power. Assuming an equal share between $\pi-$ and $\pi/2$-pulses, we use an average required power per qubit drive line of $-78\,$dBm for the estimates presented in Section \ref{sec:setupdrivelines}. We note that the duty cycle of drive pulses can be significantly lower than 33\% if the durations of two-qubit gates and readout pulses are significantly longer than the duration of drive pulses. A further reduction of the duty cycle arises if the repetition period of the algorithm is dominated by a wait time to reset the qubits to the ground state.

Readout signals used to drive readout resonators to infer the state of qubits are typically an order of magnitude smaller and have a lower duty cycle. They are therefore not taken into account in our estimates.


Concerning dissipation in flux lines, we primarily consider DC biasing currents, which are constantly applied to set the qubit frequency, in most of the cases to the so-called sweet spot frequency $\omega_q=\omega_0$, at which $\omega_q$ is to first-order insensitive to flux noise \cite{koch_charge-insensitive_2007}. Assuming a worst case scenario of random offset magnetic fields in the SQUID loops, the compensating flux $\Phi_{\rm offset}$ which we apply to reach the closest sweet spot, is equally distributed in the interval $[-\Phi_0/2,\Phi_0/2]$. This corresponds to a current interval of [-1,1]\,mA, when using a reasonable mutual inductance between flux line and SQUID loop of $M=\partial\Phi/\partial I=0.5$\,$\Phi_0/{\rm mA}$. The mutual inductance is determined by the coupling strength (proximity) of the on-chip flux line to the SQUID loop and the area of the SQUID loop. A large $M$ on the one hand minimizes the required current and thus dissipation in the line, and allows for the usage of low-noise current sources which typically have a small dynamic range. On the other hand, a large SQUID loop is also more susceptible to (external) magnetic flux noise, and care needs to be taken in the design of large coupling strengths in order to keep residual capacitive coupling of the qubit to the flux line low, which otherwise can limit the qubit lifetime due to the Purcell effect \cite{koch_charge-insensitive_2007}. The dissipated heat due to the bias currents is most critical at the MXC stage because least cooling power is available there. The amount of dissipation
depends on the DC resistances of the stainless steel cable and low-pass filter installed between CP and MXC, and to which stage this dissipated heat predominantly flows to. We experimentally determine this heat load as discussed in Section \ref{activeloadmeasFL}.

We also consider dissipation due to flux pulses, which are used for the realization of two-qubit gates. Assuming that during the flux pulse the qubit's frequency is tuned by about 10\% of its sweet spot value \cite{dicarlo_demonstration_2009,barends_superconducting_2014}, a flux amplitude of $\pm 0.2\,\Phi_0$ is needed, corresponding to a current amplitude of 0.4\,mA. The associated active loads will be estimated using the results from the DC measurements in Section \ref{activeloadmeasFL}. To estimate the duty cycle of flux pulses we consider alternating single- and two-qubit gates as discussed above, and that the flux pulse for a two-qubit gate is applied only to one of the qubits, yielding a value of $0.5\times 66\,\% = 33\,\%$.

\subsection{Radiative load}\label{sec:radiativeload}
Each temperature stage except for CP is fitted with a dedicated heat shield. For a given temperature stage $i$ we calculate the radiative heat load on heat shield $i$ from the enclosing heat shield by assuming infinitely extended cylindrical heat shields, and solving a system of coupled heat equations. The heat shields are made of Aluminium on the 50K and 4K stage and of Cu on Still and MXC. They are characterized by their emissivity $\epsilon=0.06$, as quoted by the manufacturer.
Compared to the cooling power of each stage, we find a significant contribution only for the 50K stage, amounting to $\sim 50\,$W. This corresponds to about half of the nominal cooling power of the two pulse tube coolers (2x Cryomech PT 420). This load however has already been taken into account in the available cooling power as specified by the manufacturer (40-50\,W). Radiative loads at the lower temperature stages are insignificant \cite{parma_cryostat_2015}.


\section{Cryogenic setup}\label{sec:setup}

\begin{figure}
	\includegraphics[width=\textwidth]{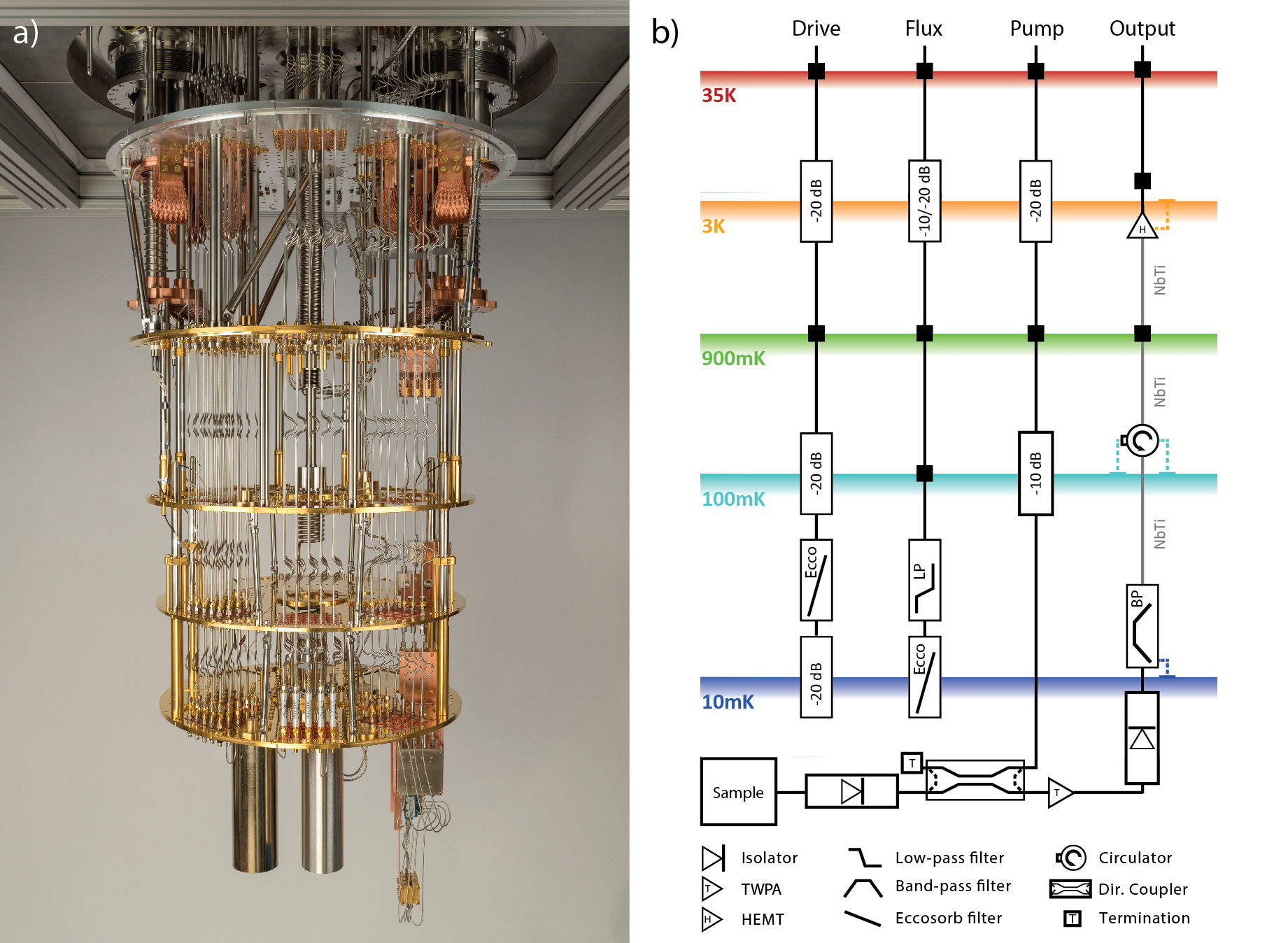}
	\caption{{\bf Cabled dilution refrigerator (DR).} a) Bluefors XLD DR with 25 drive lines, 25 flux lines, 4 read-out, 6 read-in, and 5 pump lines installed (see end of Section \ref{sec:setupdilfridge} for details). b) Schematic of the cabling configuration inside the DR.}
	\label{fig:DR}
\end{figure}

\subsection{Dilution refrigerator (DR)}\label{sec:setupdilfridge}
\begin{figure}
	\includegraphics[width=\textwidth]{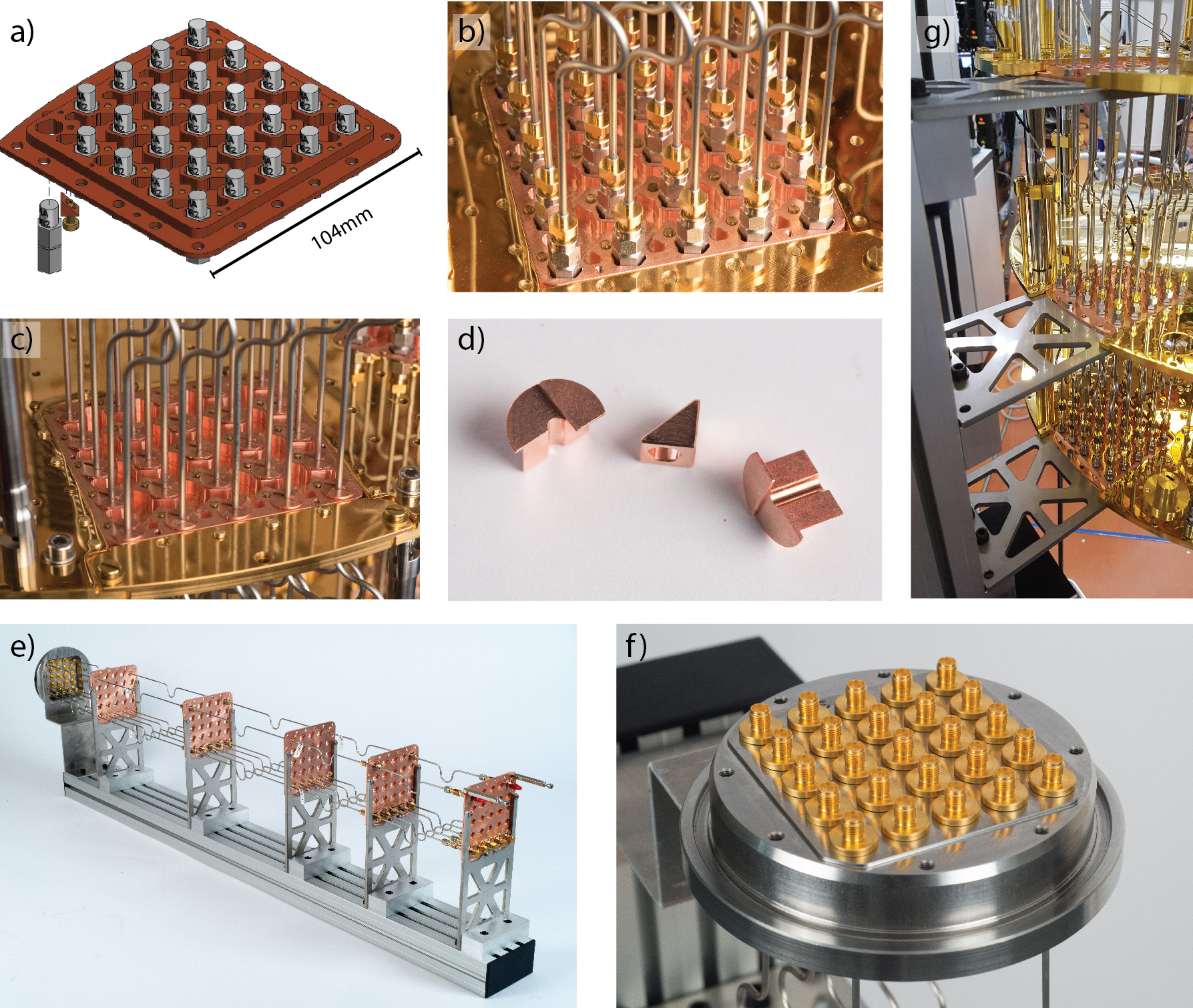}
	\caption{{\bf Thermalization concept and pre-assembly of cable trees. }a) CAD drawing of the Cu plate used to mount attenuators. To clamp the attenuators to the copper plate, an adapter as shown in d) is used. The Cu plates are then screwed onto the corresponding plate of the DR. b) Photograph of such a mounting plate with cabling installed in the DR. c) Photograph of the same plate, here used to clamp RF cables using the radiation tight Cu adapter pieces shown in d). d) Two-part Cu adapter that is pressed around an RF cable using the wedge shaped piece in the middle in combination with a screw (see also a)). e) Preassembly tool with mounted Cu plates and vacuum flange. f) Vacuum flange with vacuum tight SMA adapters installed. g) Mounting the preassembled cable tree into the DR using a custom mounting tool.}
	\label{fig:thermalizationscheme}
\end{figure}

A customized version of a Bluefors XLD400 dilution refrigerator as described in this paper is shown in Fig. \ref{fig:DR}\,a. It has a dilution unit positioned in the center of the MXC plate, making space available for an additional line-of-sight (LOS) port at the circumference of the system. A total of six LOS side ports are available in the system for the installation of RF cables. A LOS port consists of a series of vertically aligned cutouts in the plates of the dilution refrigerator designed for the installation of Cu plates for the thermalization of cables and attenuators, see Fig. \ref{fig:thermalizationscheme}\,a-c. Instead of the circular cutouts we designed rectangular cutouts with a width of 88\,mm. They allow for mounting fully assembled cable trees from the side using a custom mounting tool, see Fig. \ref{fig:thermalizationscheme}\,e-g. This has the advantage that the cable trees, comprising 25 RF lines and the vacuum flange, can be pre-assembled and tested outside of the dilution refrigerator. Two out of the eight sectors at the circumference at the 50K and 4K stages are occupied by two pulse tube coolers , but we make use of the corresponding cutouts on the Still, CP, MXC stages for installing readout components, as discussed below.

Furthermore, the distance between the Still plate and the CP has been enlarged by 60\,mm to provide more space for the inline integration of MW components, and for reducing the passive heat load due to the cables from Still to CP. The dilution refrigerator was pre-equipped by the manufacturer according to our specifications with two looms of Cu twisted pairs (AWG 35) and two looms of Ph-Br twisted pairs (AWG 36), running both from room temperature to 4K. From 4K to MXC two looms of NbTi twisted pairs were pre-installed. Each of the looms contains 12 twisted pairs.

We have measured the cooling powers available on all stages of our custom dilution refrigerator (in the presence of the pre-installed DC wiring), see Table \ref{tab:DR}, using the technique described in Section \ref{sec:refmeas}. These powers are not to be exceeded to guarantee a base temperature below 20\,mK \cite{bluefors_cooling_2014}. We designed and planned the wiring and attenuation scheme of the dilution refrigerator with the goal to not exceed those cooling powers.
The table also contains the RF cable lengths between the stages. Note that the cables have bends (bending radius $\sim 1\,$cm) to reduce strain on RF connectors when the system is cooled down.

As discussed in Section \ref{sec:diffheatloads}, good thermalization of the cables and attenuators is important for both reducing thermal noise photons affecting qubit coherence and reducing the heat load on the lower temperature stages affecting the cryostat performance. For this purpose, we have developed a flexible thermalization method for both attenuators and cables. It consists of a Cu plate with hexagonal cutouts, in which we individually clamp attenuators using a wedge shaped Cu piece, which presses the attenuator against two faces of the cutout when tightening a three millimeter (M3) Brass screw (Fig. \ref{fig:thermalizationscheme}\,a and b). The torque applied on each screw for fastening amounts to 0.9\,Nm. We use Brass screws as they thermally contract more than Cu when cooled down, providing additional clamping force to the attenuator. This is important, because in general the contact resistance between two metal pieces at low temperatures is dominated by clamping force rather than surface area \cite{salerno_thermal_1997}. We thermalize RF cables using the same Cu plate (Fig. \ref{fig:thermalizationscheme}\,c) in combination with a two-part Cu adapter (Fig. \ref{fig:thermalizationscheme}\,d). The round flange on those pieces guarantees radiation tight mounting. Using those plates we fit 25 RF lines into each of the six LOS ports.

We now continue to describe in more detail the custom cabling of the dilution refrigerator. Schematics of drive lines, flux lines, and lines for readout are shown in Fig. \ref{fig:DR}~b and are typical for superconducting qubit experiments. To minimize passive heat load, we choose UT85-SS-SS cables for all types of lines except for the output lines\footnote{Smaller diameter cables of the same type would further reduce the passive heat load, but introduce a larger frequency dependence in the S21 transmission parameter.}. We characterize and test the system at half of its capacity by installing three out of six possible cable trees in the dilution refrigerator: a cable tree comprising of 25 drive lines, a cable tree comprising of 25 flux lines, and a cable tree comprising of lines needed for readout. This readout cable tree consists of six drive lines and 5 pump lines (see Section \ref{sec:setupoutputlines}), running both from top to bottom of the dilution refrigerator, and four output lines running from room temperature to 4K. The remaining parts of the output lines continue in the non-LOS cutouts below one of the pulse tube coolers, where more space is available, see Fig. \ref{fig:outputlinecomponents}.

\subsection{Drive line configuration}\label{sec:setupdrivelines}
\begin{figure}
	\center
	\includegraphics[width=\textwidth]{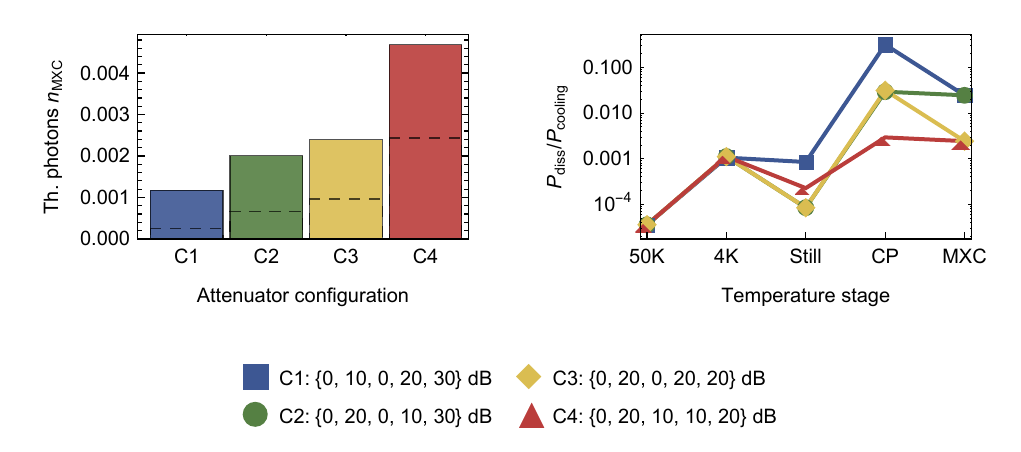}
	\caption{{\bf Thermal noise photons and signal dissipation.} a) Thermal noise photon number $n_{\rm MXC}$ at 6\,GHz for four attenuator configurations C1-C4 in the drive lines (see text). Taking into account attenuation in the cables reduces $n_{\rm MXC}$ to the values indicated by the dashed bars. b) Dissipated power on the indicated stages of the DR due to the operation of 25 drive lines with signal levels translating into an average power level of -78\,dBm at the input of the QP. Calculations for the four attenuator configurations C1-C4 are shown.}
	\label{fig:signal-dissipation}
\end{figure}

Based on our discussion of thermal noise photons in Section \ref{sec:attenuation} here we motivate our choice of attenuator configuration of 20\,dB attenuators at each of the 4K, CP, MXC stages. We first plot the thermal noise photon number at MXC $n_{\rm MXC}$ in a coaxial cable connecting the room temperature vacuum flange to MXC, for four attenuator configurations each with 60\,dB of total attenuation, see Fig. \ref{fig:signal-dissipation}\,a. The calculations assume as effective attenuator temperatures the plate temperatures listed in the second column of Table \ref{tab:DR}. For configuration C1=\{0, 10, 0, 20, 30\}\,dB we obtain $n_{\rm MXC}\approx0.0012$, which is only by 12\% larger than the lower bound $n_{\rm MXC,min}\approx0.001$ for a total of 60\,dB attenuation calculated in Section \ref{sec:attenuation}. The reason for the low $n_{\rm MXC}$ of C1 is that $n_{\rm MXC}$ considered as a function of attenuation on each of the stages is in the linear regime for configuration C1, see Fig. \ref{fig:thermal-photons}\,a. For a more precise calculation of $n_{\rm MXC}$ we take into account the attenuation in the RF cables, which sum up to about 9\,dB when combining room temperature and low temperature data \cite{kurpiers_characterizing_2017}, see Appendix \ref{app:cableattenuation}. Using a continuous version of Eq. \ref{eqn:attenuator} and assuming linear temperature gradients in the cables connecting the stages, we obtain $n_{\rm MXC}$ as indicated by the dashed bars in Fig. \ref{fig:thermal-photons}\,a. However, for simplicity in the following quantitative discussion we stick to the calculations of $n_{\rm MXC}$ for which attenuation in cables has been neglected (solid bars in Fig. \ref{fig:thermal-photons}\,a). The corresponding dissipated power in 25 drive lines on each temperature stage is normalized to the available cooling powers (Table \ref{tab:DR}), see Fig. \ref{fig:signal-dissipation}\,b. The calculations assume an average power level  of -78\,dBm required at MXC (see Section \ref{sec:requsignallevel}) and also take into account attenuation in the RF cables. While the relative loads on the 50K, 4K, and Still stages are negligibly small, the CP and MXC stages are subject to significant loads. Since configuration C1 has a particularly large load of 35\% on the CP stage, we also consider configurations with a total of 40\,dB attenuation on the CP and MXC stages instead of 50\,dB.

Removing 10\,dB of attenuation at the CP and adding 10\,dB at the 4K stage results in C2=\{0, 20, 0, 10, 30\}\,dB and reduces the active load on the CP by a factor 10. The resulting value of $n_{\rm MXC}=0.002$ is almost a factor of two larger than in C1 because 20\,dB of attenuation at the 4K stage already corresponds to the regime where $n_{\rm MXC}$ as a function of $A_{\rm 4K}$ starts to saturate to the 4K noise floor, see Fig. \ref{fig:thermal-photons}\,a. We also note that dissipation on the 4K stage is relatively low compared to the available cooling power and thus redistributing some of the attenuation at 4K to the 50K stage is not necessary. Configuration C3=\{0, 20, 0, 20, 20\}\,dB corresponds to C2 with 10\,dB of attenuation moved from the MXC to the CP stage. In this case the noise photon number only increases by 20\% compared to C2 to $n_{\rm MXC}=0.0024$. This is because at $A_{\rm CP}=20\,$dB $n_{\rm MXC}$ considered as a function of $A_{\rm CP}$ is still relatively far from reaching the CP noise floor. At the same time the relative active load on the MXC stage is reduced by a factor of 10 to $<1\%$.

To further reduce the active load on the CP by a factor ten one can redistribute 10\,dB from the CP to the Still, resulting in configuration C4=\{0, 20, 10, 10, 20\}\,dB. However, $n_{\rm MXC}$ thereby increases by a factor of two compared to C3, which can be explained with similar arguments as above. A relatively small increase in $n_{\rm MXC}$ as compared to C3 can be achieved by redistributing only 3\,dB from the CP to the Still, reducing the load on the CP by a factor two. However, this is not very practical since it comes along with a non-standard attenuation of 17\,dB at the CP. In addition, solutions with four attenuators require an additional hardware cost of one attenuator and one connectorized RF cable per line as compared to solutions with only three attenuators per line.

As a compromise between the number of noise photons and a low heat load on MXC we select C3, with a noise photon number of about 0.1\% (taking attenuation in cables into account) and a relatively low active load on MXC.

To filter infrared radiation that can lead to quasi-particle generation and thus to a reduced $T_1$ time \cite{catelani_quasiparticle_2011,barends_minimizing_2011,corcoles_protecting_2011,pop_coherent_2014,kreikebaum_optimization_2016}, we provide the option to install a custom-built IR filter based on Eccosorb CR-124 absorber material right before the attenuator at MXC, see Appendix \ref{app:Sparam}. With an attenuation of 6\,dB at 6\,GHz, it further reduces $n_{\rm MXC}$ by a factor four, while increasing dissipation on all stages except MXC by a factor four.

\subsection{Flux line configuration}\label{sec:setupfluxlines}
To reduce current and thus magnetic flux noise in the SQUID loop of a qubit, a suitable amount of attenuation needs to be installed in the flux lines. Part of this attenuation can be realized at room temperature to reduce 1/f noise and white noise from the instruments connected to the flux line (e.g. an arbitrary waveform generator) to the level of room temperature Johnson-Nyquist noise.
To reduce the noise level below room temperature Johnson-Nyquist noise we install a single attenuator $A_{\rm 4K}$at the 4K stage.
The current-noise power spectral density after this attenuator is calculated as
\begin{equation}\label{eqn:currentnoise}
S_{I}^{\rm th, eff} = \frac{S_{I}^{\rm th}(T_{\rm RT})}{A_{\rm 4K}} + \frac{A_{\rm 4K}-1}{A_{\rm 4K}}S_{I}^{\rm th}(T_{\rm 4K}),
\end{equation}
where $S_{I}^{\rm th}(T) = 2 {\rm k_B} T/R$ is the two-sided current-noise power spectral density of Johnson-Nyquist noise. The resistance $R=50\,\Omega$ corresponds to the internal impedance of a typical arbitrary waveform generator instrument.
Note that as opposed to using Eq. \ref{eqn:JNnoisequantum} in the context of thermal noise in the drive lines, considering the classical Johnson-Nyquist formula here is sufficient since the thermal energy at $T_{\rm 4K}\approx 3$\,K is much larger than the relevant noise frequencies of $\lesssim 1$\,GHz. We plot $S_{I}^{\rm th, eff}$ as a function of $A_{\rm 4K}$ for $T_{\rm RT}\approx 300$\,K and $T_{\rm 4K}\approx 3$\,K in Fig. \ref{fig:thermal-photons}\,b. Similar to Fig. \ref{fig:thermal-photons}\,a $S_{I}^{\rm th, eff}$  saturates at the thermal noise floor of the 4K stage for attenuator values $A_{\rm 4K}\gtrsim 20$\,dB.

We use $S_{I}^{\rm th, eff}$ for deriving upper bounds on the coherence times $T_2^*$ and the echo coherence time $T_2$ due to thermal noise in the flux lines \cite{ithier_decoherence_2005,koch_charge-insensitive_2007}
\begin{equation}\label{eq:T2limitwhite}
T_2^* = \frac{1}{2} T_2^E = \frac{2}{D^2 S_I}.
\end{equation}
Here, $S_I(\omega)=\int {\rm d}t\langle\delta I(0) \delta I(t)\rangle {\rm e}^{-i\omega t}$ is the power spectral density of the current noise and $D$ the sensitivity of the qubit transition frequency $\omega_{\rm ge}$ to the current $I$ in the flux line
\begin{equation}\label{eqn:partials}
D = \frac{\partial\omega_{\rm ge}}{\partial I} = \frac{\partial\omega_{\rm ge}}{\partial \Phi} \frac{\partial\Phi}{\partial I}.
\end{equation}
Using a typical mutual inductance between flux line and SQUID loop of $M=\frac\partial\Phi\partial I=0.5$\,$\Phi_0/{\rm mA}$ we obtain $T_2^*=(46, 426, 2333)\,\mu$s for an attenuation of (0, 10, 20)\,dB, at a 10\% detuning from the sweet spot frequency of a qubit with a symmetric SQUID loop \cite{hutchings_tunable_2017}. We thus conclude that an attenuation of 10-20\,dB at the 4K stage is sufficient to mitigate qubit dephasing due to thermal noise. Noise from electronic instruments connected to the flux line should be reduced to the level of Johnson-Nyquist noise by installing a suitable amount of attenuation at room temperature outside of the cryostat. Other sources of magnetic flux noise include surface adsorbates and defects \cite{kumar_origin_2016,de_graaf_direct_2017,sendelbach_complex_2009} and external magnetic field fluctuations in combination with insufficient magnetic shielding.

\subsection{Output line configuration}\label{sec:setupoutputlines}
\begin{figure}
	\includegraphics[width=\textwidth]{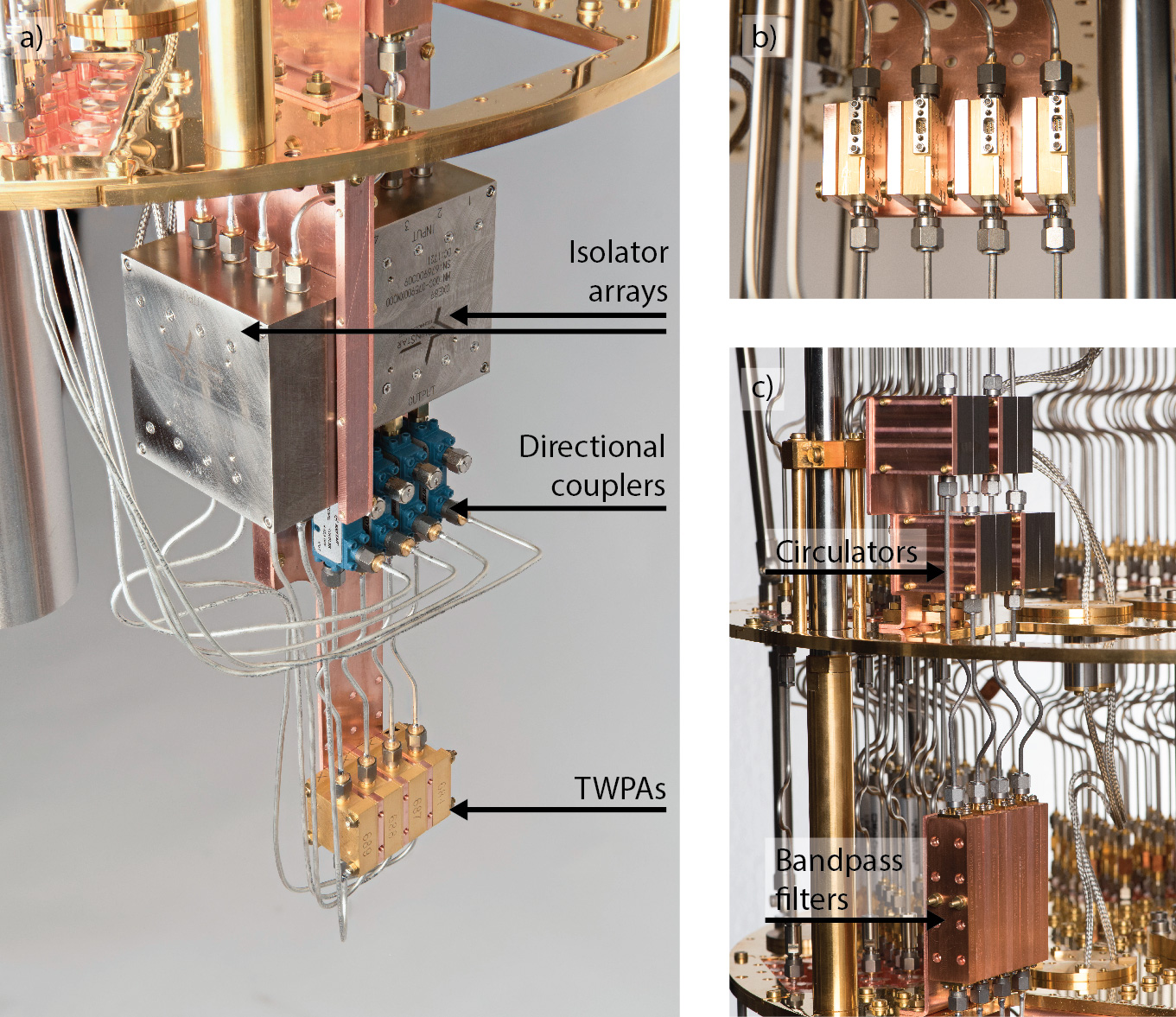}
	\caption{{\bf Integration of four high-bandwidth output lines. } a) Components mounted on the MXC plate: readout signals from the quantum processor travel through a first 4-channel, magnetically shielded isolator array (right), followed by directional couplers, TWPAs and a second, identical isolator array (left). b) Four HEMT amplifiers mounted on the 4K plate. c) Four bandpass filters, mounted and thermalized at the MXC plate, followed by four circulators mounted to the cold plate.}
	\label{fig:outputlinecomponents}
\end{figure}

Our four output lines are designed for the detection of signals on the single photon level \cite{eichler_characterizing_2012,walter_rapid_2017}. We use travelling wave Josephson parametric amplifiers (TWPA) \cite{macklin_nearquantum-limited_2015} with a broadband gain of 20-30\,dB over a frequency range of 3-12\,GHz as a near-quantum limited amplifier installed at MXC, see Fig. \ref{fig:outputlinecomponents}\,a. High-electron mobility transistor (HEMT) amplifiers (LNF LNC4\_8C) with a gain of about 40\,dB are installed at the 4K stage, see Fig. \ref{fig:outputlinecomponents}\,b. More than a total of 60\,dB isolation towards the sample is provided by a circulator (Quinstar QCY-060400CM00) at CP and two isolators at MXC. The two isolators with an operation range of 3-12\,GHz are also required for impedance matching of the TWPA over its entire gain bandwidth. For this purpose, one of the isolators is installed before and one after the TWPA, see Fig. \ref{fig:DR}\,b. Each isolator is part of a four-channel isolator array (Quinstar QCC-075900XM000), see Fig. \ref{fig:outputlinecomponents}\,a. Circulators and isolators are both magnetically shielded.

A reflective cavity bandpass filter (Keenlion KBF-4/8-2S) at MXC suppresses noise at frequencies outside the bandwidths of circulator and isolators, see Fig. \ref{fig:outputlinecomponents}\,c. To minimize signal loss between the two amplfiers, we use superconducting NbTi coaxial cables between MXC and 4K. To minimize passive heat load we chose stainless steel cables from 4K to room temperature. According to Friis formula the increased loss compared to e.g. SPCW cable does not decrease the signal to noise ratio of the amplifier chain. E.g. a loss of 10\,dB in the cable section between HEMT and the first amplifier at room temperature can be tolerated if one uses a low noise amplifier with a noise temperature of about $500$\,K as a first amplifier at room temperature\footnote{When using only a HEMT amplifier as a cryogenic amplifier (and not a TWPA), an ultra-low noise amplifier with a noise temperature of about 100\,K is required}.

For the operation of the TWPA a microwave pump signal with a typical frequency in the range 6-8\,GHz and a power level at the input of the TWPA of about $-55$\,dBm is required. This pump signal is added to the TWPA via the isolated port of a 20\,dB directional coupler\footnote{Choosing a smaller coupling ratio would decrease the transmission of the readout signal through the directional coupler accordingly.} (Krytar 120420), see Fig. \ref{fig:DR}\,b. Due to the relatively large pump power we use pump lines with a total of 50\,dB attenuation as compared to the 60\,dB in the drive lines. This 50\,dB of attenuation includes the 20\,dB attenuation provided by the directional coupler with its coupled port terminated with a 50\,$\Omega$ load resistance. Photographs of the readout components described in this Section are shown in Fig. \ref{fig:outputlinecomponents}. The TWPAs and the isolator arrays are mounted to 2\,mm thick Cu sheets whose stability is enforced by Cu bars at their borders, see Fig. \ref{fig:outputlinecomponents}\,a. For compact mounting and thermalization, the four TWPAs and the four bandpass filters (Fig. \ref{fig:outputlinecomponents}\,c) are first stacked using Cu spacers between them, which are then mounted to the Cu sheet or to an L-shaped bracket, respectively. Similarly, the four HEMT amplifiers and the four circulators are first mounted onto Cu pieces which are then mounted to a common bracket.

Using the concept of frequency-multiplexed readout \cite{heinsoo_rapid_2018}, each readout line allows for the simultaneous readout of up to 10 superconducting qubits. A pre-requisite for fast frequency-multiplexed single-shot readout is a near-quantum limited amplfier with a bandwidth of at least 1\,GHz such as a TWPA.

\section{Direct measurements of the passive heat load}\label{sec:passiveloadmeas}
To determine the passive heat load induced by each cable tree and to extract a heat load per line, we initially performed reference measurements of the cooling power in the dilution refrigerator system as delivered without microwave cabling. Then we cooled down the system each time we installed a new cable tree and recorded the temperature increases on each of the stages of the dilution refrigerator to infer the corresponding passive heat loads. The cooldown of cable trees comprising up to 25 identical lines of a given type has the advantage of generating temperature increases that are larger than typical run-to-run temperature variations on the stages of the dilution refrigerator. Furthermore, this method naturally averages over variations in material and components, as well as over possible variations occurring in the installation and thermalization of individual cables and attenuators.

\subsection{Reference measurements}\label{sec:refmeas}
\begin{figure*}
	\includegraphics[width=\textwidth]{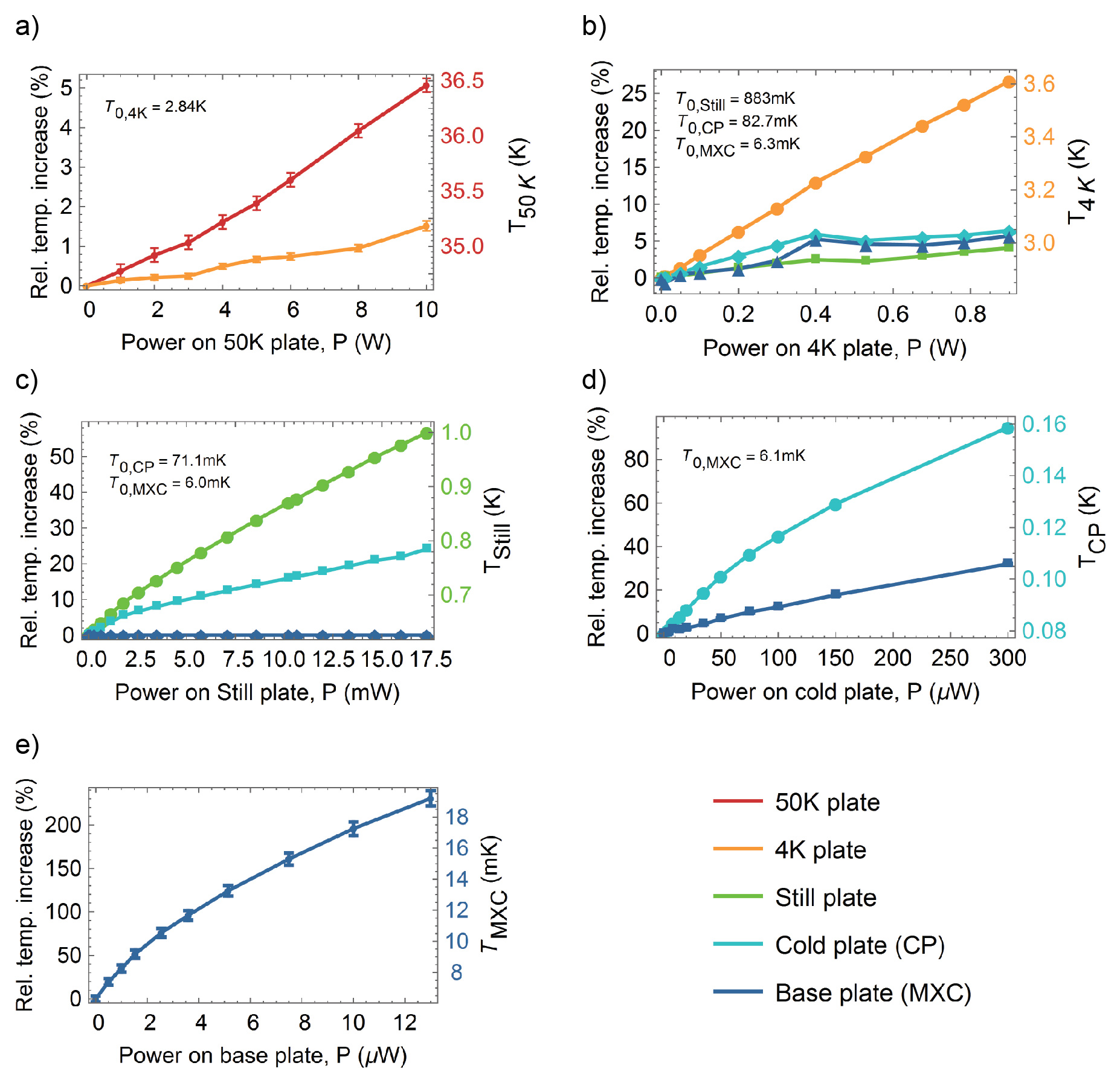}
	\caption{{\bf Reference measurements in the DR system as delivered.} Temperature increases on the different plates of the DR as a function of applied power to the 50K (a), 4K (b), Still (c), CP (d), MXC (e) stages. The temperatures indicated in the upper left corner of the plots represent the starting temperatures, without heat power applied.}
	\label{fig:passiveLoadGraphs}
\end{figure*}

Our reference measurements consist of applying a well defined heat load to a given temperature stage and recording the increased steady state temperature increases on all the other stages. In steady state the applied heat load corresponds to the cooling power on that stage at its steady state temperature. The measured relative temperature increases are shown as a function of heat load applied to a given stage in Fig. \ref{fig:passiveLoadGraphs}. The absolute temperature increase on the stage, to which the power is applied, is indicated on the right axis of each plot.
The 50K stage and the 4K stage show approximately linear temperature increases over the range of applied powers (up to 10\,W for the 50K stage and 0.9W for the 4K stage).
Heating a given stage also affects stages at lower temperature because they are thermally connected via structural elements of the dilution refrigerator such as posts and other elements such as the dilution unit and pre-installed DC wiring. In addition the $^3{\rm He}$ reaching the Still in the gas phase for condensation is effectively hotter when 50K and 4K stages are heated leading to a larger load on the Still.

The cooling power on the Still stage is provided mainly by the evaporation of $^3{\rm He}$ and reflects the dependence of the partial pressure of $^3{\rm He}$ on temperature \cite{betts_1989,bluefors_manual_2016}. The temperature of the Still plate increases by about 60\% over the range of applied heat powers of 0-18\,mW, see Fig. \ref{fig:passiveLoadGraphs}\,c. Importantly, the heat power applied to the Still stage also sets the cooling power on MXC, $P_{\rm MXC}\propto \dot{n}_3T_{\rm MXC}^2$ \cite{betts_1989} because it regulates the flow $\dot{n}_3$ of the $^3{\rm He}$ through the dilution unit and thus across the phase boundary between the concentrated and dilute phases of $^3{\rm He}$ in the mixing chamber. Since an increased Still temperature also creates an additional load on the CP and MXC stages, there is an optimum flow value which maximizes the cooling power for a given $T_{\rm MXC}$. The $T_{\rm MXC}^2$ dependence of $P_{\rm MXC}$ arises because the enthalpy difference between the two phases is proportional to $T_{\rm MXC}$ \cite{betts_1989}. For all measurements we set $P_{\rm Still}=10.7$\,mW, corresponding to a flow of  0.69(1)\,mmol/s and a cooling power on MXC of about 5\,$\mu$W (13\,$\mu$W) at 13\,mK (20\,mK), see Fig. \ref{fig:passiveLoadGraphs}\,e. Importantly, when increasing the flow to 1.0\,mmol/s by applying a heat power of 40\,mW to the Still we reach a cooling power of 19\,$\mu$W (540\,$\mu$W) at 20\,mK (100\,mK).

The CP has no active cooling element, but since it is mounted on top of the last set of heat exchangers \cite{bluefors_manual_2016} it is effectively cooled by the cold mixture entering and leaving the mixing chamber. We measure the temperature increase over a relatively wide range of heat powers up to 300\,$\mu$W since our attenuator configuration C3 in the drive lines is expected to create a significant active load on the CP.
Over this range the temperature of the CP doubles from about 80\,mK to about 160\,mK, see Fig. \ref{fig:passiveLoadGraphs}\,d. The effect on $T_{\rm MXC}$ is relatively low, e.g. at the largest measured value of $P_{\rm CP}=300\,\mu$W $T_{\rm MXC}$ rises from 6.1\,mK to 8.1\,mK, corresponding to an effective load of only 1\,$\mu$W on the MXC stage.

\subsection{Determination of passive heat loads}\label{sec:detPassiveLoads}
The reference measurments presented in Fig. \ref{fig:passiveLoadGraphs} allow us to convert an increase in temperature $\Delta T_i$ observed on stage $i$ after the cooldown of a given cable tree, into a passive heat load
\begin{equation}\label{eqn:passiveloadextr}
\Delta P_i = \frac{\partial P_i}{\partial T_i}\Delta T_i
\end{equation}
on that stage. The coefficients $\frac{\partial P_i}{\partial T_i}$ are extracted from a linear fit to the low (relative) temperature regimes of the data in Fig. \ref{fig:passiveLoadGraphs}. All measured $\Delta T_i$ are sufficiently small to justify the use of this linear expansion. Additional loads due to an increased temperature of the next higher stage $i-1$ are subtracted by substituting $\Delta T_i\rightarrow \Delta \tilde{T}_i=\Delta T_i-\frac{\partial T_i}{\partial P_{i-1}}\Delta P_{i-1}$ in the r.h.s. of Eq. \ref{eqn:passiveloadextr}. This corresponds to a shift of the data in Fig. \ref{fig:passiveLoadGraphs} along the horizontal axes in negative direction. We have verified this procedure by measuring $T_i$ vs. $P_i$ curves for $P_{i-1}\neq 0$. The resulting data sets lie on top of their $P_{i-1}= 0$ counterparts in Fig. \ref{fig:passiveLoadGraphs} when shifting the data along the applied power axis by an amount corresponding to $(\partial T_i/\partial P_{i-1}) P_{i-1}$. A significant correction $>20$\% results only for $\Delta T_{\rm Still}$.

Additional care needs to be taken when determining $\Delta T_{\rm CP}$ and $\Delta T_{\rm MXC}$. This is because an increased Still temperature changes the flow through the dilution unit thereby not only leading to an additional passive load on CP and MXC, but effectively changing the cooling power vs temperature curves. This is most evident for the MXC stage when considering the relation $P_{\rm MXC}\propto \dot{n}_3T_{\rm MXC}^2$ from above. Before determining $\Delta T_{\rm CP}$ and $\Delta T_{\rm MXC}$ we thus reduce the Still heat power such that we reach $T_{\rm Still}=882(1)\,$mK, correponding to the flow of 0.69(1)\,mmol/s set for the reference measurements.

\renewcommand{\arraystretch}{2.6}
\begin{table}
	\centering
	\resizebox{\textwidth}{!}{%
		\begin{tabular}{cl|rrrrr}
			\hline
			& \multicolumn{1}{l|}{} & \multicolumn{1}{r}{\textbf{50K}} & \multicolumn{1}{r}{\textbf{4K}} & \multicolumn{1}{r}{\textbf{Still}} & \multicolumn{1}{r}{\textbf{CP}}& \multicolumn{1}{r}{\textbf{MXC}}\\
			\hline

			\multirow{2}{*}{\rotatebox[origin=c]{90}{\parbox[c]{1.7 cm}{\centering Drive line UT085 SS-SS}}}   & \textbf{Measured HL} & 45(23)\,mW & 1.0(2)\,mW & 4(3)\,$\mu$W & 0.4(2)$\mu$W & 0.013(6)$\mu$W\\
			        & \textbf{Predicted HL} & 24-27\,mW & 0.4-1.7\,mW & 1.9-2.1\,$\mu$W & 0.33-0.43\,$\mu$W & 0.004\,$\mu$W\\
			\hline

			\multirow{2}{*}{\rotatebox[origin=c]{90}{\parbox[c]{1.7cm}{\centering Flux line UT085 SS-SS}}} 	& \textbf{Measured HL} & 56(27)\,mW & 1.2(2)\,mW & 2(1)\,$\mu$W & 0.29(3)\,$\mu$W & 0.025(5)\,$\mu$W\\
			& \textbf{Predicted HL} & 24-27\,mW & 0.4-1.7\,mW & 1.9-2.1\,$\mu$W & 0.30\,$\mu$W & 0.027-0.131\,$\mu$W\\
			\hline

			\multirow{2}{*}{\rotatebox[origin=c]{90}{\parbox[c]{1.7cm}{\centering Output line UT085 NbTi}}}
			& \textbf{Measured HL} & - & - & - & 0.29(12)\,$\mu$W & 0.020(16)\,$\mu$W\\
			& \textbf{Predicted HL} &-&-&-&  0.29-0.39\,$\mu$W &  0.012-0.023\,$\mu$W\\
			\hline
		\end{tabular}
	}
	\caption{{\bf Passive heat load per line. } Passive heat load (HL) of cable types installed in the DR, as inferred from observed temperature increases after the installation of new cables into the DR. The upper and middle sections of the table refer to 0.085" diameter stainless steel coaxial cable (UT085 SS-SS) with attenuator configurations used in drive lines and flux lines, respectively. Error bars include statistical errors between different cooldowns and systematic uncertainties in the temperature measurements. The intervals of predicted heat loads correspond to calculations of lower and upper bounds (see text).}
\label{tab:passiveload}
\end{table}

To extract the passive heat loads for each type of line we separately cooled down the dilution refrigerator after installation of the drive line cable tree, the readout cable tree without the NbTi lines, the four NbTi output lines between the 4K stage and the MXC stage, and the flux line cable tree, respectively. From the extracted total passive heat load after each cooldown we calculate an average load per line by dividing the total loads by the number of lines in the cable tree.

To extract a mean passive load per drive line we combine the data from the cooldown of the drive line tree and the readout tree since the readout tree consists of drive lines and pump lines having the same thermalization scheme with attenuators at the 4K, CP, and MXC stages. The data is shown in the first section of Table \ref{tab:passiveload}.
The determined heat loads are in qualitative agreement with predictions based on Eq. \ref{eqn:heat-flow}, see row "Predicted HL" in the upper part of Table \ref{tab:passiveload}. The lower and upper bounds of the intervals represent estimates in which thermalization of the center conductor and dielectric on stages without attenuator (50K and Still stage) is fully achieved or not at all. In calculating the predicted passive load we have assumed that the center conductor is fully thermalized on stages with an attenuator. However, it is an open question how well the center conductor thermalizes to the CC inside an attenuator and it is likely to depend on the physical structure and thermal properties of the attenuator \cite{yeh_microwave_2017}.

While the measured loads on the 50K, 4K, Still, CP stages agree within error bars with the predicted values, the load on the MXC stage is larger than predicted. This can have several reasons.
First, there could be discrepancies between the thermal conductivitiy of the particular stainless steel (SS) of the cables and the data used for the estimates in that temperature interval. Second, thermalization of the center conductor via the attenuator at the CP could be ineffective. Third, the long timescales for thermalization of components at MXC, in particular if they contain hydrogen \cite{Schwark1983}, can lead to an overestimation of the heat load since we measure $\Delta T_i$ always at the same time ($\sim$1\,day after the condensation of the mixture when the MXC temperature is sufficiently stable). Fourth, gas describing from newly installed components' outgassing can freeze out at MXC, creating an additional heat load.

The average passive loads per flux line on the 50K, 4K, and Still stages agree within error bars with the corresponding passive loads per drive line, see second section of Table \ref{tab:passiveload}. This is expected since the two types of line are nominally identical from the vacuum flange to the Still stage. The load on the CP is also comparable to the corresponding load per drive line. This observation is also expected because the additional load on the CP in case of full thermalization of the center conductor as compared to no thermalization at all is only by 0.03\,$\mu$W (0.13\,$\mu$W) larger if the closest upper stage at which the center conductor is thermalized is the Still (4K) stage. Note that this corresponds to the interval of 0.33-0.43\,$\mu$W for the predicted load per drive line as compared to the predicted load per flux line of 0.3\,$\mu$W. The intervals for the predicted loads are calculated for complete thermalization of the center conductor at the 50K and Still stages or no thermalization of the center conductor at all.
The passive load per flux line on MXC is significantly larger than the one per drive line. This is due to the absence of an attenuator in the flux line at the CP causing a direct passive load from the Still to the MXC stage via the center conductor. Thermalization of the center conductor at the Still stage is likely because otherwise a load of 0.131\,$\mu$W at MXC is expected, which is by a factor of four larger than what we measure.

The passive loads extracted from the cooldown of the four NbTi lines are listed in the third section of Table \ref{tab:passiveload}, showing agreement with the predictions. Since it is unclear whether a circulator helps for achieving thermalization of the center conductor we calculate lower and upper bounds for the predicted heat loads by assuming complete thermalization of the center conductor at Still and CP or no thermalization at all. However, the error bars on the measured values are too large to draw any conclusion in that respect.

In summary, after the installation of 66 RF lines, mainly UT085-SS-SS, $T_{\rm MXC}$ increased from 6.1(1)\,mK to 8.4(2)\,mK, corresponding to a total passive heat load on MXC of $1.4(2)\,\mu$W. This is more than a factor of ten smaller than the available cooling power at 20\,mK, and thus sets the stage for experiments with tens to hundreds of superconducting qubits.\\

\section{Active load measurements}\label{sec:activeloadmeas}

\subsection{Drive lines}

\begin{figure}
	\center
	\includegraphics{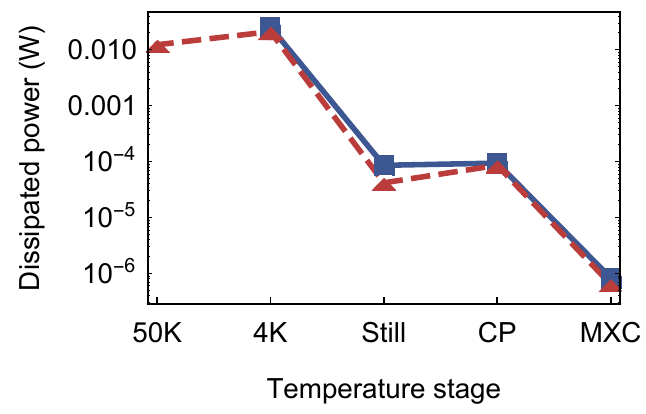}
	\caption{{\bf Dissipation in drive lines.} Dissipated power at the temperature stages of the DR when applying a microwave tone at 5\,GHz and a power of 16\,dBm, corresponding to a signal level of -52.5\,dBm at MXC. Shown are measured data (blue rectangles, solid line) and results of calculations (red triangles, dashed line).}
	\label{fig:dissipation-driveline}
\end{figure}

To experimentally test loads due to signal dissipation we apply a RF signal at one of the drive lines, and measure the corresponding temperature increases on the stages in the dilution refrigerator. In particular, we attempt to evaluate whether the power dissipated in an attenuator predominantly flows to the stage on which the attenuator is thermalized, or whether part of that heat flows to the next lower stage. This is important because the signal levels at stages with attenuators differ by 20\,dB in the selected attenuator configuration of C3=\{0, 20, 0, 20, 20\}\,dB. In other words, the power dissipated in an attenuator at the 4K (CP) stage is by a factor 100 larger than the power dissipated in the attenuator at the CP (MXC) stage. Hence, if only a small fraction of heat dissipated in an attenuator at a given stage flows towards the lower temperature stages, this can significantly increase the heat load on these stages.

We first apply a power of 10\,dBm at a frequency of 10\,MHz, at which dissipation in the RF cables themselves is negligible. This allows for a simple comparison with the expected loads of 10\,mW on the 4K stage, 100\,$\mu$W on CP, and 1\,$\mu$W on MXC. Using our reference measurements, we extract an active load of $11.5(1.6)\,$mW on the 4K stage, 95(9)\,$\mu$W on CP, 1.1(1)\,$\mu$W on MXC, in agreement with the predicted values and showing that there is no significant extra load on these stages due to signal dissipation on higher stages. However, we measured a non-zero active load of 48(6)\,$\mu$W on the Still stage, which we attribute to a small fraction of the dissipated power at 4K (about 0.4\%) flowing towards the Still. We note that the extracted active loads represent pure active loads, i.e. we subtract from the measured load on a given stage the additional passive load due to an increased temperature of the next higher stage.

We next apply a signal at 5\,GHz with a power of 16\,dBm such that the resulting temperature increases on the lower stages are well measurable. To calculate the power level at MXC, we have to take into account the attenuation in the cables \cite{kurpiers_characterizing_2017}, see Appendix \ref{app:cableattenuation}.
The total attenuation in the cables sums up to about 8.5\,dB, and with the 60\,dB of discrete attenuation yields a signal level at MXC of about -52.5\,dBm, much larger than the average value of -78\,dBm required in a drive line, see Section \ref{sec:attenuation}. To predict an active load per drive line the extracted pure active loads shown as blue rectangles in Fig. \ref{fig:dissipation-driveline} need to be scaled down accordingly. On the stages with attenuators the data agrees well with the estimated active loads (red triangles). In our calculations we attribute one half of the dissipated power in a cable section itself (excluding the attenuators) to the stage above and the other half to the stage below. On the Still stage the measured load is by a factor two larger than estimated, which is consistent with the above observation made at a signal frequency of 10\,MHz.

The active load per TWPA pump line is found to be comparable to the active load per drive line, with a dominating contribution of 0.23\,$\mu$W on the CP stage. This is a result of a pump power of about -65\,dB required at the input of the TWPA and the pump line having 10\,dB less attenuation than the drive lines.

Summarizing, we have experimentally verified the predicted active loads for the selected attenuator configuration in the drive lines as presented in Fig. \ref{fig:signal-dissipation}\,b. Since this configuration has been chosen to keep the active load significantly below the available cooling powers, the active load due to execution of single-qubit gates has a negligible influence on the temperature of the quantum processor. Whether the thermal photon number is as low as predicted needs to be measured \cite{yan_flux_2016,yeh_microwave_2017} and would provide information about the effective temperatures at which the attenuators emit blackbody radiation.

\subsection{Flux lines}\label{activeloadmeasFL}

\begin{figure}
	\center
	\includegraphics[width=0.45\textwidth]{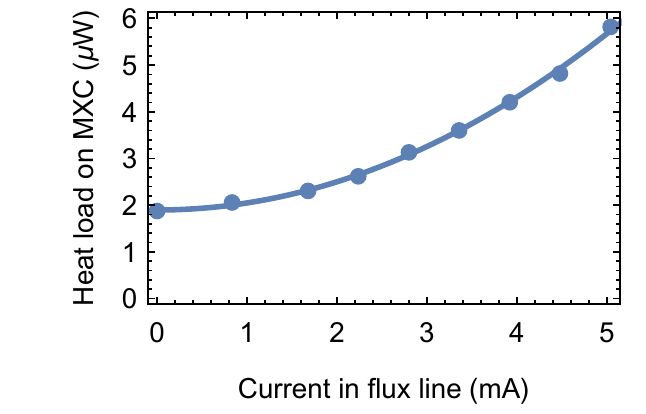}
    \hspace{0.5cm}
    \includegraphics[width=0.45\textwidth]{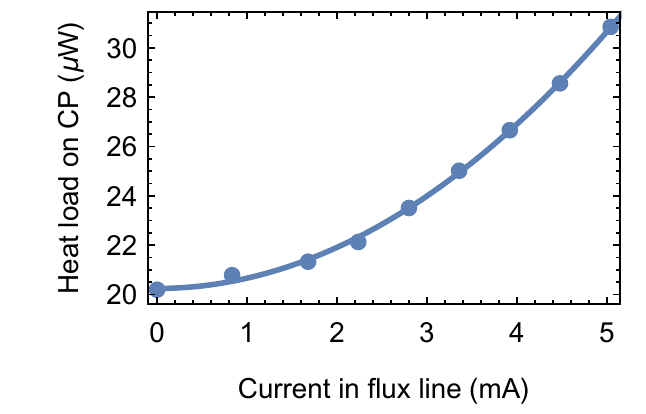}
	\caption{{\bf Dissipation in flux lines.} Dissipated power at MXC (left) and CP (right) measured as a function of current in the flux line. Solid lines are quadratic fits to the data.}
	\label{fig:dissipation-fluxline}
\end{figure}

To determine the active load arising from flux biasing the qubits, we terminate one of the flux lines at MXC with an electric short, apply a DC current to it, and measure the corresponding temperature increase and heat loads. While the relative temperature increases are negligibly small on the stages from room temperature to the Still, they are significant on CP and MXC. As expected we observe a quadratic increase of the extracted loads $P_{\rm MXC/CP}=R_{\rm eff,MXC/CP}I_{\rm MXC}^2$ as a function of the applied current. The maximum applied current amounted to 8.8\,mA at the input of the dilution refrigerator or to 4.6\,mA after the 10\,dB attenuator at 4K, and thus at MXC (denoted $I_{\rm MXC}$). From quadratic fits to the data we extract $R_{\rm eff,MXC}=0.15(1)\,{\rm \Omega} = 0.15(1)\mu{\rm W/mA^2}$ and $R_{\rm eff,CP}=0.42(2)\,{\rm \Omega}$.

The sum of the two resistances corresponds to the resistance of a coaxial stainless steel cable with a length of 20\,cm, which is a bit smaller than the cable length of 29\,cm from Still to MXC. How the dissipated power in that cable distributes to the Still, CP, and MXC stages depends on details such as the thermal conductances between the center conductor, which contributes most to the resistance of the coaxial cable, and all of the stages.


For a reasonable choice of mutual inductance between flux line and qubit discussed in Section \ref{sec:requsignallevel}, the DC biasing currents needed to tune the qubits to the desired transition frequencies are considered to be randomly distributed in the interval $[-I_{\rm max},I_{\rm max}]=[-1,1]$\,mA. Hence, an average power of $P_{\rm avg}=\frac{1}{I_{\rm max}}\int_0^{I_{\rm max}} R_{\rm eff} I_{\rm MXC}^2\,{\rm d}I_{\rm MXC} = \frac{1}{3} R_{\rm eff} I_{\rm max}^2 \approx 0.050(3)\,\mu$W is dissipated. For 25 flux lines this sums up to 1.25(7)\,$\mu$W, which is about a factor of 15 smaller than the cooling power available at MXC at a temperature of 20\,mK. We note that this is a worst-case scenario since with careful magnetic shielding it could be possible to reduce the offset fluxes to close to zero.

For an estimate of dissipation due to flux pulses used for two-qubit gates we use the measured effective DC resistances, the required current amplitude of about 0.2\,mA and a duty cycle of 33\% (see Section \ref{sec:requsignallevel}). The resulting heat loads on CP and MXC are less than 20\% of the corresponding loads due to DC flux biasing, see Fig. \ref{fig:totalheatbudget}.

\section{Conclusion and Outlook}\label{sec:conclusion}

\begin{figure}
	\center
	\includegraphics{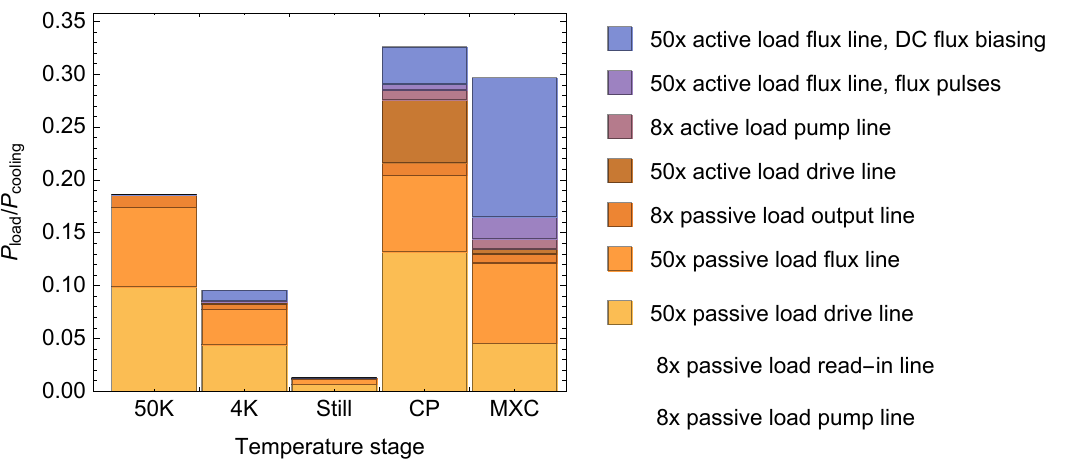}
	\caption{{\bf Total heat load budget for the operation of a 50 qubit processor.} Predicted passive and active heat loads for the operation of a 50 qubit processor with individual drive and flux control. The heat loads are normalized to the measured cooling powers as presented in Table \ref{tab:DR}.}
	\label{fig:totalheatbudget}
\end{figure}

Based on our measurements of passive and active loads we predict a total heat load on the dilution refrigerator when operating a 50 qubit processor with individual drive and flux control and with a multiplexed readout architecture allowing for simultaneous readout of sets of 6-7 qubits, see Fig. \ref{fig:totalheatbudget}. Such a quantum compressor requires a total of 124 RF lines (50 drive lines, 50 flux lines, 8 output lines, 8 read-in lines, 8 TWPA pump lines), corresponding to the operation of our system at its full capacity. The largest relative loads of about 30\,\% occur at the CP and MXC stages, with about an equal share between passive and active loads. The load on MXC corresponds to an operation temperature of 14\,mK. The active load in the drive lines is a result of targeting a noise photon number of about $10^{-3}$. The relative loads of about 20\,\% and 10\,\% on the 50K and 4K stages, respectively, are dominated by passive loads. The load on the Still stage is negligibly small. Hence, the thermal performance of our system would allow for the operation of a 150 qubit processor if the capacity for coaxial line integration was increased three-fold. This could be achieved either by increasing the density of coaxial line integration or by building a larger diameter dilution refrigerator.


The material of the coaxial cables has been chosen to minimize passive heat load on all of the stages. However, instead of the standard 0.085\," diameter coaxial cables, one could use 0.047\," cables or even thinner. This is beneficial because the passive heat load scales with the square of the diameter, whereas the attenuation in the cable scales approximately linearly. For the proposed case of 0.047\," SS-SS cables, the passive heat load would be reduced by almost a factor of four. Since large attenuation is desired in the drive lines, the active load due to the execution of single-qubit gates would not be increased if one reduced the attenuation of the installed attenuators accordingly. However, the dominant active load at MXC due to flux biasing of the qubits would be increased four-fold because the DC resistance of the cables scales inversely with their diameter.

We briefly discuss an optimistic scenario in which magnetic flux offsets are brought close to zero by careful magnetic shielding of the quantum processor. Disregarding thus active load due to  DC flux biasing and considering the use of 0.047\," diameter cables, the dominant relative load on the CP stage amounts to about 13\,\%, allowing for the operation of about 400 qubits.

Ultimately, we propose the following modifications to allow for the operation of about thousand qubits. First, we use superconducting NbTi cables for the flux lines eliminating active load in these lines. Second, we tolerate a CP temperature of about 200\,mK at which a cooling power of at least 400\,$\mu$W is available. To maintain the same noise photon number of 0.001 we shift about 7\,dB of attenuation from the CP to the MXC stage, increasing the active load on MXC by a factor six. Third, we tolerate a base plate temperature of 30\,mK at which the cooling power at MXC is at least doubled to about 40\,$\mu$W, while not affecting the noise photon number. We note that in this optimized scenario active loads in the drive lines are the dominating heat load at CP and MXC.

We conclude with the remark that future large-scale dilution refrigerators will likely have several dilution units to provide sufficient cooling power at the CP and MXC stages.

\appendix*
\section*{Appendices}

\section{Attenuation of different types of coaxial cables}\label{app:cableattenuation}
\begin{figure*}
	\center
	\includegraphics[width=0.7\textwidth]{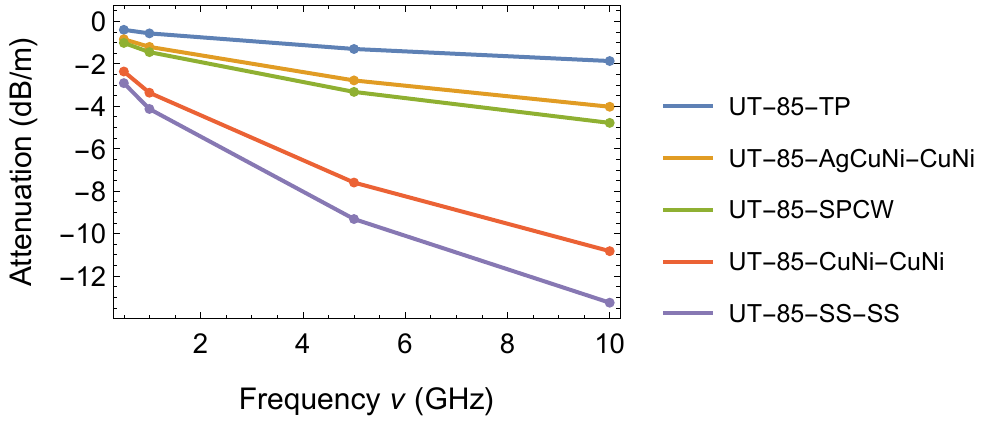}
	\caption{{\bf Attenuation in coaxial cables at room temperature.} Attenuation in indicated types of coaxial cables as a function of frequency as specified by the manufacturer \cite{microcoax,COAX}.}
	\label{fig:cableAtt}
\end{figure*}

The attenuation at room temperature of various types of coaxial cables is shown as a function of frequency in Fig. \ref{fig:cableAtt}. At low temperature the attenuation is slightly reduced, e.g. the attenuation constant of a UT-85-SS-SS cable at 6\,GHz reduces from 9.7\,dB/m at room temperature to 8.2\,dB/m at 4\,K \cite{kurpiers_characterizing_2017}. For the calculation of the thermal noise photon number and the active loads presented in Fig. \ref{fig:signal-dissipation} and in Fig. \ref{fig:dissipation-driveline} we have taken such temperature dependences to first order into account.

\section{Thermal conductivities of relevant materials and cables}\label{app:thcond}
\begin{figure}
	\center
	\includegraphics[width=0.7\textwidth]{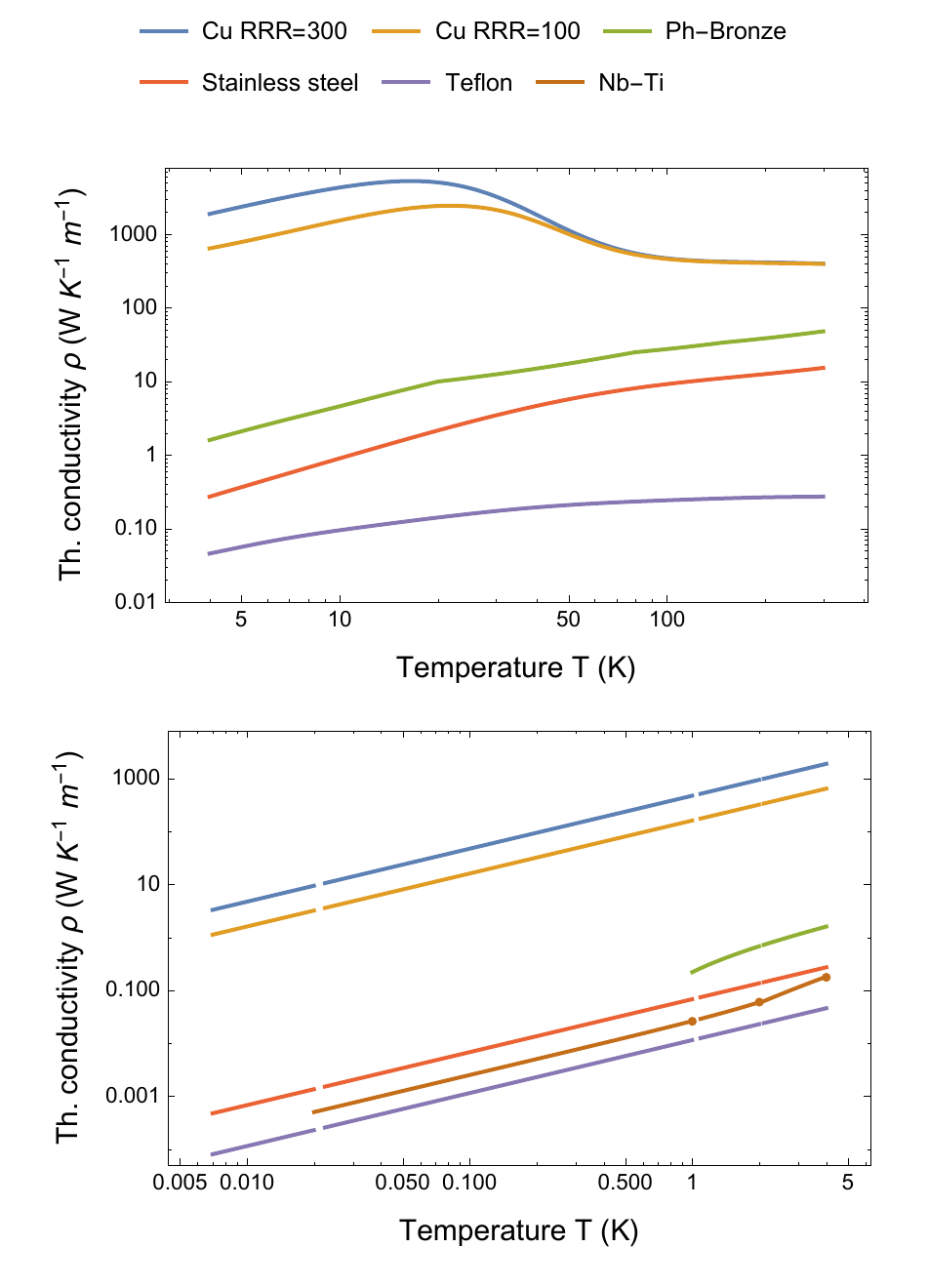}
	\caption{{\bf Thermal conductivities.} Thermal conductivity of indicated materials as a function of temperature \cite{nist_cryoMaterials_2000,lakeshore_phbr_2018} on a logarithmic scale. The upper plot shows the temperature range from 4K to 300K, the lower plot shows the range from 0.01K to 4K. Data for NbTi in the high temperature regime is taken from \cite{flachbart_thermal_1978}. The low temperature curve of NbTi (lower plot) is created from three data points \cite{cern_materialProperties_2014} and a linear extrapolation to zero, see text.}
	\label{fig:thermalCond}
\end{figure}

The thermal conductivities of typical cryogenic cable materials are shown as a function of temperature in Fig. \ref{fig:thermalCond}. The data has been used for the estimation of passive heat loads. Data for most of the materials is only available down to $4\,$K. From 4\,K to 0\,K we use a linear extrapolation to zero. This is justified for metals at low temperatures, where the electronic contribution to the thermal conductivity dominates \cite{cern_materialProperties_2014}. Although the thermal conductivities of crystallized thermal insulators such as PTFE, and also of superconductors, is expected to have a $T^3$ dependence at low temperature, we make a linear interpolation to zero as an upper bound for our estimates.

\section{S-parameters of RF lines}\label{app:Sparam}
\begin{figure*}
	\includegraphics[width=\textwidth]{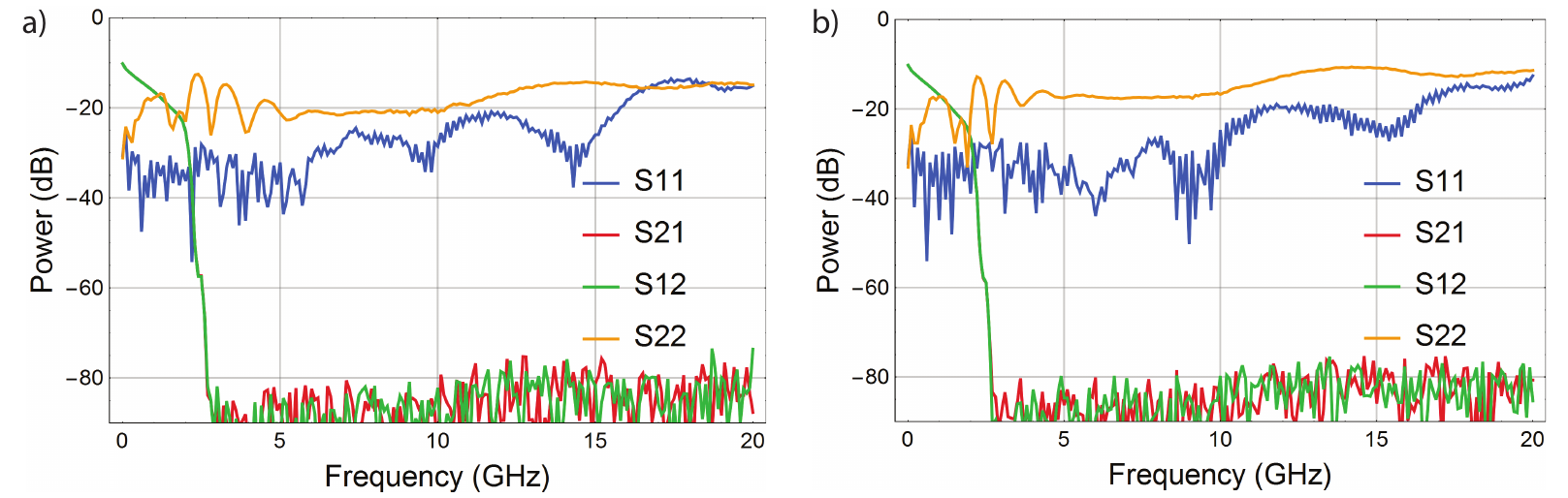}
	\caption{{\bf Typical S-parameters of installed RF lines.} a) Example of a drive line with a total of 60\,dB attenuation and an Eccosorb filter having an attenuation of $\sim 5\,$dB at 5\,GHz. b) Example of a flux line with 10\,dB attenuation, a low pass filter (Minicircuits VLFX1350), and an Eccosorb filter with an attenuation of $\sim 3\,$dB at 1\,GHz $(\sim 20\,$dB at 5\,GHz).}
	\label{fig:S_parameters}
\end{figure*}

Typical S-parameter response curves of the installed drive lines and flux lines are shown in Fig. \ref{fig:S_parameters}\,a. Each line has been measured from the port on the vacuum flange (Port 1) to the port at the MXC flange (Port 2), using a vector network analyzer (VNA). The drive lines typically achieve reflection parameters below -20dB from 0 to 12\,GHz, and below -25dB in the range of 4-8\,GHz, which is most relevant for the applied microwave pulses. For lines with an Eccosorb filter we choose to position it before the attenuator at MXC such that the line is well impedance matched also at Port 2 even if the Eccosorb filter itself is not perfectly impedance matched.

The transmission coefficients confirm that the signal loss at DC frequencies is set by the installed attenuators (3x 20dB). For higher frequencies, the transmission coefficients decrease as expected due to the frequency dependent signal loss in the cables and Eccosorb filter.

The transmission spectrum of the flux line starts at -10\,dB, corresponding to the 10\,dB attenuator installed at the 4K stage, and sharply drops at $2\,$GHz due to a low pass filter with a cutoff frequency of 1.3\,GHz, see Fig. \ref{fig:S_parameters}\,b. The continuous drop between 0 and 2\,GHz is due to a combination of frequency dependent loss in cable and Eccosorb filter. The reflection S22 from the sample side towards the Eccosorb filter is below -20\,dB in the range of 0-500\,MHz. Since this range is larger than the typical bandwidth of a flux pulse, possible multiple reflections of flux pulses between the sample and the Eccosorb filter are significantly suppressed.

\section{Temperature sensors and heaters}
\label{app:tempsensorsheaters}
We use a PT100 temperature sensor at 50K, Cernox CX-1010 sensors at 4K and Still, and RuO2 RX-102B sensors at CP and MXC (all sensors are from Lake Shore Cryotronics). All except the sensor at CP were pre-installed by the manufacturer. The sensors are read out using the temperature controller model 372 from Lake Shore. We also installed resistive heaters on stages that had no heater installed by the manufacturer, i.e. on the 50K and the CP stage. On the 50K stage we installed a 25\,$\Omega$, 100\,W cartridge heater (HTR-50, Lakeshore)  We use resistive cartridge heaters from Lakeshore Cryogenics. On the 50K stage we installed a 25\,$\Omega$, 100\,W heater (HTR-25-100), whereas on the CP we installed a  50\,$\Omega$, 50\,W heater (HTR-50). For mounting, the cartridge heaters are clamped in a circular hole in a rectangular Cu piece that is then mounted to the corresponding plate in the cryostat.

\section*{Acknowledgements}
The authors thank Y. Salath\'e, G. Norris, C. K. Andersen, and S. Gasparinetti for useful discussions, and M. Frey for support during the installation of the dilution refrigerator. This work is supported by the Office of the Director of National Intelligence (ODNI), Intelligence Advanced Research Projects Activity (IARPA), via the U.S. Army Research Office grant W911NF-16-1-0071 and by ETH Zurich. The views and conclusions contained herein are those of the authors and should not be interpreted as necessarily representing the official policies or endorsements, either expressed or implied, of the ODNI, IARPA, or the U.S. Government. The U.S. Government is authorized to reproduce and distribute reprints for Governmental purposes notwithstanding any copyright annotation thereon.

\bibliographystyle{naturemag}
\bibliography{biblio}

\end{document}